\documentclass[aps,pre,longbibliography,twocolumn,eqsecnum,letterpaper,10pt]{revtex4-2}

\usepackage{graphicx}% Include figure files
\usepackage{dcolumn}% Align table columns on decimal point
\usepackage{bm}% bold math

\usepackage[utf8]{inputenc}
\usepackage[T1]{fontenc}
\usepackage{etoolbox}
\usepackage{float}
\usepackage[usenames,dvipsnames]{xcolor}
\usepackage{cancel}

\usepackage{comment}
\usepackage{mathtools}
\usepackage{amsmath,amssymb}
\usepackage{hyperref}% add hypertext capabilities
\newcommand{\bg}{{\bf g}}
\newcommand{\bq}{{\bf q}}
\newcommand{\br}{{\bf r}}
\newcommand{\bk}{{\bf k}}
\newcommand{\al}{\alpha}
\newcommand{\be}{\beta}

\newcommand{\mL}{\mathcal{L}}
\newcommand{\mH}{\mathcal{H}}

\newcommand{\mP}{\mathcal{P}}
\newcommand{\mQ}{\mathcal{Q}}

\newlength\figurewidth
\AtBeginDocument{\setlength\figurewidth{.5\linewidth}}

\def\kT{\ensuremath{k_\text{B}T}}

\newcommand{\ghostequal}{\mathrel{\phantom{=}}} %space taken by "="

\newcommand{\ii}{\mathrm i}
\newcommand{\ee}{\mathrm e}
\newcommand{\dd}{\mathrm d}

\newcommand{\liou}{\mathcal{L}}
\newcommand{\opq}{\mathcal{Q}}

\newcommand{\scal}[1]{\langle #1 \rangle}

\newcommand{\thermodevin}[3]{\frac{\partial #1}{\partial #2}\big|_{#3}} % "in text", smaller "|"
\newcommand{\da}{\delta a}
\newcommand{\db}{\delta b}

\newcommand{\de}{\delta e}

\newcommand{\dmu}{\delta \mu}

\newcommand{\daag}{\da_\mathbf g}
\newcommand{\dagp}{\da_{\mathbf g'}}

\newcommand{\dn}{\delta n}
\newcommand{\dng}{\delta n_\mathbf g}

\newcommand{\ddp}{\delta p}

\newcommand{\ds}{\delta s}

\newcommand{\dya}{\delta y_\alpha}

\newcommand{\dT}{\delta T}

\renewcommand{\ng}{n_\mathbf g} % normally used for the phonetic "ng"
\newcommand{\ngp}{n_{\mathbf g'}}

\newcommand{\bgp}{{\mathbf g'}}
\newcommand{\bj}{\mathbf j}

\newcommand{\bu}{\mathbf u}
\newcommand{\bv}{\mathbf v}

\newcommand{\ga}{g_\alpha}
\newcommand{\gb}{g_\beta}

\newcommand{\ja}{j_\alpha}
\newcommand{\jb}{j_\beta}

\newcommand{\mub}{\mu_\beta}

\newcommand{\qa}{q_\alpha}
\newcommand{\qb}{q_\beta}
\newcommand{\qc}{q_\gamma}

\newcommand{\ra}{r_\alpha}
\newcommand{\rb}{r_\beta}

\newcommand{\ua}{u_\alpha}
\newcommand{\ub}{u_\beta}
\newcommand{\uc}{u_\gamma}

\newcommand{\dua}{\delta \ua}
\newcommand{\dub}{\delta \ub}
\newcommand{\duc}{\delta \uc}

\newcommand{\ddua}{\delta \dot u_\alpha} % not the same as \dot \ua !!
\newcommand{\ddub}{\delta \dot u_\beta}

\newcommand{\va}{v_\alpha}

\newcommand{\vd}{v_\delta}

\newcommand{\hab}{h_\albe}

\newcommand{\sigab}{\sigma_\albe}

\newcommand{\sigcd}{\sigma_\gade}

\newcommand{\albe}{{\alpha\beta}}

\newcommand{\beal}{{\beta\alpha}}

\newcommand{\bega}{{\beta\gamma}}

\newcommand{\gade}{{\gamma\delta}}

\begin{document}

\preprint{APS/123-QED}

\title{Continuum mechanics for the elastic properties of crystals: 
Microscopic approach based on projection-operator formalism}
% Force line breaks with \\
%alternative title:
%\title{Hydrodynamic equations of non-ideal crystals: Derivation starting from microscopic density fluctuations}
%\title{Hydrodynamics of crystals: Microscopic approach based on projection operator formalism and density functional theory}
\author{Florian Miserez}
\author{Saswati Ganguly}
\author{Rudolf Haussmann}
\author{Matthias Fuchs}
\affiliation{Fachbereich Physik, Universit\"at Konstanz, D-78457 Konstanz, Germany}

\date{\today}% It is always \today, today,

% \begin{abstract}
% Interpreting and predicting macroscopic measurable properties like elastic constants, heat conductance etc from underlying microscopic interactions and correlations of ordered solids have remained an active area of research. Despite the relative ease of framing the problem, this has remained a surprisingly difficult challenge in the field of classical statistical mechanics. Measurable material properties at macroscopic length scales obey the laws of thermodynamics which subsumes the huge number of microscopic events occurring at several orders of magnitude smaller length and time scales. Continuum level descriptions, deriving constitutive relations like the relation connecting stress and strain in a mechanical deformation experiment, often rely on phenomenological parameters or models. In this paper we provide a microscopic basis to the phenomenological hydrodynamic description of a crystalline solid with local defects.
% \end{abstract}

\begin{abstract}
We present a microscopic derivation of the laws of continuum mechanics of nonideal ordered solids including dissipation, defect diffusion, and heat transport. Starting point is the classical many-body Hamiltonian.  The approach relies on the 
Zwanzig-Mori projection operator formalism to connect microscopic fluctuations to thermodynamic derivatives and transport coefficients. Conservation laws and spontaneous symmetry breaking, implemented via Bogoliubov's inequality,  determine the selection of the slow variables. Density fluctuations in reciprocal space encode the displacement field and the defect concentration.   Isothermal and adiabatic elastic
constants are obtained from equilibrium correlations, while transport coefficients are given as Green-Kubo formulae, providing the basis for their measurement in atomistic simulations or colloidal experiments. The approach and results are compared to others from the literature.
\end{abstract}

\maketitle

%\tableofcontents

\section{\label{intro}Introduction}
Continuum mechanics describe the physical properties of condensed matter at large spatial and temporal scales. For simple liquids, the theory is called hydrodynamics where there are five relevant variables for the five physical degrees of freedom : the mass density $\rho_m$, three components of the momentum density $j_x$, $j_y$, $j_z$, and the energy density $e$. In crystalline solids,  spontaneous symmetry breaking takes place and a periodic structure arises for all physical microscopic variables. Therefore, on macroscopic scales, three more relevant variables arise and these are the three components of the displacement field $u_x$, $u_y$, and $u_z$. These additional physical degrees of freedom means there are a total of eight relevant variables for a crystalline solid.

However, in conventional elasticity theory there are only seven degrees of freedom \cite{LL_ET}. In case of defect-free ideal crystals with perfect lattice structures, each lattice site is occupied by exactly one particle. This constrains the density to be the divergence of the displacement field ${\bf u}$. In this way the physical degrees of freedom are reduced by one, from eight to seven.

On the other hand, a one-component crystalline solid at a non-zero temperature is expected to have a finite concentration of point-defects. These point defects may be vacant lattice sites or particles at interstitial places.  As a consequence, there is an eighth physical degree of freedom in nonideal crystals best described by the defect density $c$. This eighth degree of freedom, describing the diffusion of point-defects, has been introduced by Martin, Parodi and Pershan \cite{Martin_Parodi_Pershan}. Later, Fleming and Cohen \cite{Fleming_Cohen} further developed and elaborated a continuum-mechanics description based on the phenomenology of this idea.

The first microscopic approach, to understand reversible mechanical response and dissipative transport in crystals with local-defects, has been theorised by Szamel and Ernst \cite{Szamel_Ernst_1993,Szamel_1997}. They investigate the microscopic density $n_\mathbf{g}(\mathbf{q})$ in Fourier representation where the wave vector $\mathbf{k} = \mathbf{g} + \mathbf{q}$ is decomposed into the discrete reciprocal lattice vector $\mathbf{g}$ and the continuous wave vector $\mathbf{q}$ restricted to the first Brillouin zone. Szamel and Ernst \cite{Szamel_Ernst_1993} suggest a microscopic formula for the displacement field $\mathbf{u} = u_x \mathbf{e}_x + u_y \mathbf{e}_y + u_z \mathbf{e}_z$ associated with the linear elastic response due to small deformations of the crystal. From microscopic principles, they derive the dynamic equations for the eight degrees of freedom governing the continuum mechanics of the system. As a result they obtain explicit expressions for the linear elastic constants. Later, Szamel's \cite{Szamel_1997} extension of the theory includes dissipative effects. He applies the concepts of the projection-operator formalism \cite{Forster} and derives the Green-Kubo~\cite{Green_1,Green_2,Kubo} relations for the transport coefficients.

The microscopic approach of Szamel and Ernst \cite{Szamel_Ernst_1993} has been extended by Walz and Fuchs \cite{Walz2010}. Their representation of the microscopic particle density $n_\mathbf{g}(\mathbf{q})$ in terms of the displacement field $\mathbf{u}$ and the defect density $c$ also identifies previously ignored correlations between displacements and defect density fields. These general theoretical frameworks were first implemented to a specific model of cluster crystals by H\"aring \emph{et al.}\ \cite{Haring2015}. Cluster crystals are defect-rich crystals, where an inhomogeneously distributed number of soft particles occupies lattice sites \cite{Mladek2006,DNA_CC}. An extensive examination of the elastic properties of the same cluster crystals, taken up by Ganguly \emph{et al.}\ \cite{SG_jcp_2022}, infers how local disorder quantitatively impacts mechanical response. The perspective for the elasticity of hard-sphere crystals, presented by Lin \emph{et al.}\ \cite{lin2021} highlights the influence of microscopic interactions and direct correlations on the thermo-mechanics.  Ras \emph{et al.}\ \cite{Ras2020}, through their study of disordered binary crystals, further extends the scope of these theories. An extension to include nonlinear effects and fluctuations has been provided by Haussmann \cite{haussmann}. This theory uses projection operators defined~\cite{Kawasaki1973} for ensembles far from equilibrium but with an assumption of local entropy maximisation. Finally, an alternative and equivalent approach has been presented by Mabillard and Gaspard \cite{Mabi_2,Mabi_1}. Their approach, unlike~\cite{Szamel_Ernst_1993,Szamel_1997,Walz2010,Haring2015,SG_jcp_2022,lin2021,Ras2020,haussmann}, avoids the explicit use of projection operators.

In this paper we build on the microscopic theories of Walz-Fuchs \cite{Walz2010} and H\"aring \emph{et al.}\ \cite{Haring2015}. These previous theories are restricted to the reversible isothermal case where the temperature is constant and dissipative processes are neglected. In this paper we consider non-constant temperatures which imply heat transport by  diffusion. Furthermore, we consider dissipative contributions to calculate the transport coefficients of diffusive processes and attenuation in propagative modes. We focus on general concepts and the derivation of explicit formulas from microscopic principles.

The paper is organised as follows:
Section~\ref{theory} is devoted to developing and explaining the microscopic basis of our theoretical framework. In it, section~\ref{theory_part1} briefly summarises the conceptual premise of the Mori-Zwanzig projection operators and introduces the set of microscopic dynamical variables relevant for this paper. Section~\ref{theory_part2}, then derives the equations of motion for these relevant variables after defining the static and dynamic correlations responsible for the reactive and dissipative couplings in the equations. The main focus of section~\ref{micro_macro_fields} is to obtain the coarse-grained fields of elasticity from the microscopic fluctuating fields through an appropriate ansatz. Once the equations of motion are obtained in the reduced space of the coarse-grained relevant variables, section~\ref{thermodynamics} derives their connection to macroscopic thermodynamic properties of the system. This is achieved by the consideration of thermodynamic identities and expansions which allow us to interpret our theoretical perspective in the broader context of material properties in different thermodynamic ensembles. Finally, in section~\ref{conclusion} we conclude giving outlines of future directions.

%\section{System of interest : Face centered cubic cluster crystal}

\section{\label{theory}The microscopic theory}
\subsection{Densities of the relevant variables and their projected dynamics}\label{theory_part1}
The thermodynamics of macroscopic systems predominantly depend on the dynamics of a few relevant variables. Specialised projection tools allow the derivation of the dynamics of the few relevant variables from the microscopic degrees of freedom governed by the Hamiltonian dynamics. In the crystalline phase, the reversible parts of the equations of motion of the slowly relaxing {\it relevant variables} govern the macroscopic mechanical response, while the dissipative parts describe the coefficients associated with heat or momentum transport. The Mori-Zwanzig projection operator formalism~\cite{Zwanzig2001,Forster} provides a way to connect these measurable equilibrium and non-equilibrium thermodynamic properties to the underlying large number of microscopic degrees of freedom. The success of this formalism leading to an useful representation of a material phase relies on a good choice of the relevant variables. They will be called 'hydrodynamic variables' in the following and will be identified in the 'hydrodynamic limit' of small frequencies and long wavelengths. Conservation laws and long-ranged elastic correlations arising from spontaneous symmetry breaking provide the basis for their selection.

The equations of motion for a chosen set of relevant variables $\{A_{i}(t)\}$ within a linear response framework is the eventual outcome of the Mori-Zwanzig formalism~\cite{Zwanzig2001,Forster}. Averages $ \langle\delta \hat{A}_{k}(t)\rangle^{\text{lr}}$ of interest for time $t> 0$ describe the relaxation of small initial perturbations $ \langle\delta \hat{A}_{k}(t=0)\rangle^{\text{lr}}$ of the relevant variables generated by external fields in the past ($t<0$).  The general form of these equations and the expressions for the matrices involved are summarised here.
\begin{align}\label{EOM_gen1}
    \partial_{t}\langle\delta \hat{A}_{k}(t)\rangle^{\text{lr}}=&i\sum_{i,j}\chi^{-1}_{ij}\omega_{jk}\langle\delta \hat{A}_{i}(t)\rangle^{\text{lr}} \nonumber\\
    -&\sum_{i,j}\int_{0}^{t}d\tau \chi^{-1}_{ij}m_{jk}(t-\tau)\langle\delta \hat{A}_{i}(\tau)\rangle^{\text{lr}} +F_{k}(t)
\end{align}
The projection operation splits the time evolution of the relevant variables into reversible (first term in Eq.~\eqref{EOM_gen1}) and dissipative (second term in Eq.~\eqref{EOM_gen1}) parts. Besides these two contributions arising from the present and earlier values of the relevant variables, there exists a random force term $F_k(t)$.  As we aim to derive linear elasticity theory along with the attenuation in the elastic waves due to the dissipative couplings, only the motion of the averaged fields are of interest. Including the fluctuating forces would lead to stochastic equations introduced by Langevin. Here that would correspond to fluctuating elasticity theory which is not our aim as it would provide a far too detailed description. So we neglect the fluctuating force $F_{k}(t)$ in our calculations.  The angular brackets in Eq.~\eqref{EOM_gen1} denotes ensemble averages and the relation between the small fluctuation of the relevant variable and the average correlation functions within the linear response theory is highlighted by the use of index $``\text{lr}"$. Because of the closeness to equilibrium, the dynamical parameters entering Eq.~\eqref{EOM_gen1} can be then evaluated by (grand) canonical averaging. Yet, before defining the matrices of static susceptibility $\chi_{ij}$, frequency $\omega_{ij}$, and memory $m_{ij}$, that appear in the dynamical equations of the averaged fields, one needs to define the Liouville operator $\mL$ governing the dynamics of the microscopic variables. The Liouville operator $\mL$, acting on the dynamical variables of a system with a conserved phase space volume, is defined as the Poisson bracket of an arbitrary dynamical variable $\hat{A}$ and the Hamiltonian $\mathcal H$ of the system. The time evolution of a microscopic variable $\hat{A}$ is given by~\cite{Evans2008}
\begin{subequations}\label{def_L}
\begin{align}
    &\partial_{t}\hat{A}(t)=\{\hat{A}(t),\mH\}=i\mL \hat{A}(t)\\
    &\hat{A}(t)=e^{i\mL t}\hat{A}(0).
\end{align}
\end{subequations}
With the Liouville operator, the definitions of the matrices in Eq.~\eqref{EOM_gen1} can be given:
\begin{subequations}\label{EOM_matrix1}
\begin{align}
    &\chi_{ij}=\beta\langle \delta \hat{A}^{*}_{i}\delta \hat{A}_{j}\rangle\\
    &\omega_{jk}=\beta\langle \delta \hat{A}^{*}_{j}\mL \delta \hat{A}_{k}\rangle\\
    &m_{jk}(t)=\beta\langle \delta \hat{A}^{*}_{j}\mL \mQ e^{-i\mQ\mL\mQ t}\mQ\mL \delta \hat{A}_{k}\rangle.
\end{align}
\end{subequations}
Here, $\beta^{-1}=k_{B}T$ $i.e.$ the temperature $T$ multiplied to the Boltzmann constant $k_{B}$, represents the scale of energy in the system. As $k_{B}$ is a constant, its value is set to one without any loss of generality. The generalised Langevin equation (Eq.~\eqref{EOM_gen1}) is derived~\cite{Zwanzig2001} by splitting the Liouville dynamics onto two orthogonal subspaces described by the projectors
\begin{subequations}\label{project}
\begin{align}
    &\mP= \sum_{i,j}\; \delta \hat{A}_{i}\langle \delta \hat{A}^{*}_{i}\delta \hat{A}_{j}\rangle^{-1} \langle \delta \hat{A}^{*}_{j}=\beta \sum_{i,j}\; \delta \hat{A}_{i}\rangle\chi^{-1}_{ij}\langle\delta \hat{A}^{*}_{j}\\
    &\mQ=1-\mP
\end{align}
\end{subequations}
The latter projector $\mQ$ enters the matrix of memory functions $m_{ij}(t)$.

Guided by hydrodynamic description of fluids or magnetic systems~\cite{Forster,chaikin_lubensky_1995}, conserved quantities and variables associated with spontaneously broken continuous symmetries are chosen as the hydrodynamic variables for an ordered solid. This includes non-ideal crystals that possess finite concentrations of point-defects. Mass, three components of linear momentum and energy are the conserved variables in the system. The fluctuations in the densities of the following set of relevant variables 
\begin{align}\label{rel_var_micro}
\{\delta (\hat{A}_{i}/V)\}=\left (\delta \hat{\rho}_{\bg}(\bq), \delta \hat{j}_{\al=1,2,3}(\bq),\delta \hat{e}(\bq)\right)    
\end{align}
are chosen to describe the mechanical response and transport processes in a three dimensional crystalline solid. Here, $\bq$ is a wave vector restricted to lie in the first Brillouin zone of reciprocal space, and the limit $\bq\to0$ is of interest for deriving hydrodynamic theories. For a thermodynamic system of volume $V$, the first set of variables $\delta \hat{\rho}_{\bg}(\bq)$ are the density fluctuations with almost the periodicity of the lattice; see their definition in Eq.~\eqref{def} below. They are indexed by the reciprocal lattice vectors $\bg$, whose number we call $N$. While $N\to \infty$ is the relevant limit, considering $N$ finite helps in interpreting some algebraic manipulations later on.  The finite values  $\bg\neq 0$ reflect the broken translational symmetry of an ordered phase. The relation of $\delta \hat{\rho}_{\bg}(\bq)$ to the microscopic particle density (Eq.~\eqref{FT_rho_op}) justifies its use as the relevant variable associated with the conservation of mass as well. The second set $\delta \hat{j}_{\al=1,2,3}$ and the third variable $\delta \hat{e}$ are the fluctuations in the three components of linear-momentum-densities and the energy density. The respective conservation laws are given in Eq.~\eqref{ceq_j} and Eq.~\eqref{ceq_e}.

All the conservation laws in this section have the general form of the continuity equation
\begin{align}\label{cont_gen}
    \partial_{t}\hat{\rho}^{A}(\bq,t)+iq_{\al}\hat{j}^{A}_{\al}(\bq,t)=0
\end{align}
for the density $\hat{\rho}^{A}$ and current $\hat{j}^{A}$ of a conserved variable $A$. Here, the equations are given in reciprocal space. In real space, these equations state that for a conserved quantity, like the total energy of the system, any small temporal change in its density at some spatial point $\br$, will be due to a resultant current of that quantity to or from a small volume element around that spatial point. We denote microscopic operators with a {\it hat} on the symbol of the variable to distinguish it from the averaged quantity of the same variable (for example see Eq.~\eqref{hv_j}). The use of microscopic operators further implies that these conservation laws are obeyed locally for any microscopically defined phase space function. Macroscopic conservation laws relating the average densities to averaged currents, in the complex dynamical system, can be obtained through ensemble or coarse-grained averages. This is the conceptual basis for deriving the generalised Langevin equations (Eq.~\eqref{EOM_gen1}).

With particles of unit mass, the microscopic particle density operator is $\hat{\rho}(\br,t)=\sum_{i=1}^{N_{p}}\delta\left(\br-\br_{i}(t)\right)$ where the sum runs over all $N_{p}$ particles, indexed by $i$, in the system. In the Fourier space, it is
\begin{align}\label{FT_rho_op}
    \hat{\rho}(\bk,t)&=\int \dd^{3}r e^{-i\bk\cdot\br} \hat{\rho}(\br,t)\nonumber\\
    &=\sum_{i=1}^{N_{p}}e^{-i{\bf k} \cdot \br_{i}(t)}=\sum_{i=1}^{N_{p}}e^{-i(\bg+\bq) \cdot\br_{i}(t)}.
\end{align}
The total wave vector $\bk=\bg+\bq$ is written as a sum of the reciprocal lattice vector $\bg$ and the wave vector $\bq$ of the first Brillouin zone. 
This separation is possible because, in equilibrium crystals and in the thermodynamic limit, the averages of $\hat\rho(\bg+\bq,t)$ have only contributions at reciprocal lattice vectors and thus the ensemble averaged Bragg peak amplitudes at $\bg$ are
\begin{equation}\label{def_ng}
    n_{\bg}=\dfrac{1}{V}\langle\hat{\rho}(\bg,t)\rangle=\dfrac{1}{V}\left\langle \sum_{i=1}^{N_{p}}e^{-i\bg \cdot \br_{i}(t)}\right\rangle.
\end{equation}
The reciprocal lattice vectors $\bg$ form a Bravais lattice \cite{Ashcroft_Mermin1976}. In the absence of thermal fluctuations for an ideal crystal, with rigidly fixed particles at each lattice site, $n_{\bg}$ simplifies to the inverse of the volume of the lattice unit cells. But, in any other scenario, the deviations in $n_{\bg}$ results from particle motions due to thermal fluctuations or defect diffusion.
In ordered structures like crystalline solids, the Bogoliubov inequality~\cite{Forster,Wagner} indicates a long range correlation of the density fluctuations $\delta\hat{\rho}_{\bg}(\bq)$, %and a slow relaxation of density correlations (
whose correlation function diverges as $\propto q^{-2}$ for wave vectors close to all non-zero reciprocal lattice vectors $\bg\neq0$. This is the argument for the inclusion of the density fluctuation close to a reciprocal lattice vector, $\delta\hat{\rho}_{\bg}(\bq)$,
%Bragg peak amplitudes $n_{\bg\neq 0}$ (denoted by ``$n$'' as opposed to ``$\hat{\rho}$'' for a  general microscopic density) associated with the spontaneous breaking of symmetry in  crystals 
in the set of slow variables (Eq.~\eqref{rel_var_micro}). With the continuity equation for the mass density or number density for particles of unit mass 
\begin{eqnarray}\label{ceq_ng}
\partial_{t}\hat{\rho}(\bg+\bq,t)+i(g+q)_{\al}\hat{j}_{\al}(\bg+\bq,t)=0,
\end{eqnarray}
the amplitude at $\bg=0$ still serves as the slow hydrodynamic variable associated with the conservation of mass.

In Eq.~\eqref{ceq_ng} and subsequent equations the Einstein convention of summation over repeated indices is used. Here, $\hat{j}_{\al}$ is a component of the density of the linear momentum. The Fourier representation of the space and time dependent number densities, allows us to circumvent the necessity of defining the fields of elasticity attached to a reference lattice structure. Instead, the microscopic operator corresponding to the fluctuations in the density 
close to Bragg peaks
\begin{eqnarray}\label{def}
&\delta \hat{\rho}_{\bg}(\bq,t)=\hat{\rho}(\bg+\bq,t)-n_{\bg}V\delta_{\bq 0}
\end{eqnarray}
are defined as deviation from the equilibrium ensemble averaged Bragg peak amplitudes $n_{\bg}$ (Eq.~\eqref{def_ng}) for reciprocal lattice vectors $\bg$. Now averaging $\delta\hat\rho_{\bg}(\bq,t)$ over the linear response many-body distribution yields the observable microscopic field for the density fluctuations measured within the linear hydrodynamic regime\cite{Walz2010} when evaluated in the small $\bq$ limit
\begin{equation}\label{hv_ng}
\delta n_{\bg}(\bq,t)=\langle \delta \hat{\rho}_{\bg}(\bq,t)\rangle^\text{lr} %= \delta \rho_{\bg}(\bq,t).   
\end{equation}
Here we want to reiterate the distinct notations used for microscopic operators and averaged variables in Eq.~\eqref{hv_ng}. In all our calculations, we use symbols with {\it hat} for microscopic density operators like $\hat{\rho}_{\bg}(\bq,t)$ to distinguish them from their averaged counterparts $\langle\hat{\rho}_{\bg}(\bq,t)\rangle^\text{lr}$.
%denoted by $\rho_{\bg}(\bq,t)$ for ease of representation in their equations of motion (see Eq.~\eqref{EOM_gen1} and Eq.~\eqref{EOM_gen2}). 
For notational clarity, % similar to Eq.~\eqref{def_ng}, 
we choose to represent all average fluctuations by latin letters and thus use the symbol $\delta n_{\bg}(\bq,t)$ instead of $\langle\hat{\rho}_{\bg}(\bq,t)\rangle^\text{lr}$
%$\delta \rho_{\bg}(\bq,t)$ 
(this connects to the notation in Eq.~\eqref{def_ng}). These conventions are consistently followed for all variable notations in this paper.

The next set of slow, relevant fluctuations are given by
\begin{equation}\label{hv_j}
    \delta j_{\al}(\bq,t)=\langle \hat{j}_{\alpha}(\bq,t)\rangle^\text{lr}.
\end{equation}
justified by the conservation of linear momentum 
\begin{eqnarray}\label{ceq_j}
\partial_{t}\hat{j}_{\al}(\bq,t)+iq_{\beta}\hat{\sigma}_{\alpha\beta}(\bq,t)=0 ,
\end{eqnarray}
derived from the operator for the momentum density components $\hat{j}_{\al}(\br,t)=\sum_{i=1}^{N_{p}}p_{\alpha}(\br_{i})\delta\left(\br -\br_{i}(t)\right)$ and their corresponding spatial Fourier transforms $\hat{j}_{\al}(\bq,t)=\int d^{3}re^{-i\bq\cdot\br}\hat{j}_{\al}(\br,t)$.
The term $\hat{\sigma}_{\alpha\beta}$ is the stress tensor; see Appendix~\ref{def_micro_app} for its definition. The conservation of energy
\begin{eqnarray}\label{ceq_e}
\partial_{t}\hat{e}(\bq,t)+i\bq \cdot {\bf \hat{j}}^{e}(\bq,t)=0
\end{eqnarray}
in an isolated system also provides a relevant variable and its coupling with number density is related to heat transport and thermal expansion in the system. The spatial Fourier transform for the microscopic energy density 
\begin{subequations}\label{en_micro}
\begin{align}
\hat{e}(r,t)&=\sum_{i=1}^{N_{p}}E(\br_{i})\delta (\br-\br_{i}(t))\\
&=\sum_{i}\dfrac{\hat{j}_{\al}(\br_{i})\hat{j}_{\al}(\br_{i})}{2m}\delta(\br-\br_{i})+\dfrac{1}{2}\sum_{i\neq j}V(r_{ij})\delta (\br-\br_{i}),
\end{align}
\end{subequations}
is given by 
\begin{align}\label{FT_e_op}
    \hat{e}(\bq,t)=\int d^{3}r e^{-i\bq\cdot \br} \hat{e}(\br,t)
\end{align}
%\com{$E$ should be given}
similar to the mass and momentum densities. The microscopic definition of the energy current $\hat{{\bf j}}^{e}$ is given in the appendix~\ref{def_micro_app}. Following arguments similar to the momentum conservation law, the relevant fluctuations for the energy is denoted by 
\begin{eqnarray}\label{hv_e}
\delta e(\bq,t)=\langle \delta \hat{e}(\bq,t)\rangle^{lr}
\end{eqnarray}
This concludes the introduction of the slow variables enlisted in Eq.~\eqref{rel_var_micro} and required in providing a microscopic basis for the hydrodynamic equations of a crystalline solid including all dissipative processes such as heat and defect diffusion.

\subsection{Microscopic basis for the equations of motion in the hydrodynamic limit}\label{theory_part2}

In order to understand the dynamics of the relevant variables, that impact the macroscopic properties like the elastic constants and the different transport coefficients, we first need to focus on the microscopic definitions of the three key quantities $\chi_{ij}, \omega_{ij}$ and $m_{ij}$ (Eq.~\eqref{EOM_matrix1}) of the generalised Langevin equation (Eq.~\eqref{EOM_gen1}). In the following paragraphs~\ref{sus_micro},\ref{freq_micro},\ref{mem_micro}, we define and discuss each of these quantities. Finally, in paragraph~\ref{TE_micro} we present the microscopic time evolution equations for the  set of $N+4$ relevant variables.

\subsubsection{The static susceptibility matrix and intensive variables}\label{sus_micro}
The static susceptibility matrix contains the direct and the cross correlations, measured at equilibrium, between the different relevant variables (see Eq.~\eqref{EOM_matrix1}a). In the linear response framework, the equilibrium susceptibility quantifies the small change in a system property, for example density, on being subjected to an external field, for example a changing chemical potential. Reading this relation in the opposite direction,  intensive variables $\tilde {A_{i}}$  conjugate to the selected slow variables can be introduced.
The set of these conjugate variables $\{\tilde {A_{i}}\}$ is built from the general form 
\begin{equation}\label{conj_MZ_1}
\langle \delta \tilde{A}_{i}\rangle^{\text{lr}} = \sum_j \chi^{-1}_{ij}\langle\delta \hat{A}_{j}\rangle^{\text{lr}}.
\end{equation}
The generalised Langevin equation in Eq.~\eqref{EOM_gen1} translates to a simpler one  written in terms of these conjugate variables; see Eq.~\eqref{EOM_gen2} below. Equation~\eqref{conj_MZ_1} also helps in interpreting $\chi_{ij}$. The comparison with phenomenological approaches, e.g.~the one in Ref.~\cite{Fleming_Cohen}, can take place on the level of equations combining conjugate variable sets. Moreover, the intensive variables are used in nonlinear projection operator formalism extending our linear response study \cite{haussmann}. Therefore, we use conjugate sets of variables in the following presentation and derive explicit expressions for their equations of motion. 
Starting with the relation between the three sets of relevant variables (see Eq.~\eqref{rel_var_micro}, Eq.~\eqref{hv_ng}, Eq.~\eqref{hv_j}, Eq.~\eqref{hv_e}), their respective thermodynamic conjugates are defined using the static susceptibility matrix,
\begin{align}\label{rel_conjugates}
    &\begin{bmatrix}
    \delta\mathbf{a}_{\bg}(\bq,t)  \\
       \delta b(\bq,t)\\
    \delta \mathbf{v} (\bq,t) \\  
\end{bmatrix}
    =V\mathbf{\chi^{-1}}
\begin{bmatrix}
   \delta \mathbf{n_{\bg}}(\bq,t)\\
      \delta e(\bq,t)\\
   \delta \mathbf{j} (\bq,t) \;
\end{bmatrix}.
\end{align}
 Here $\mathbf{n_{\bg}}(\bq,t)$ and its conjugate $\mathbf{a}_{\bg}$ denote N-dimensional column vectors whose components $ n_{\bg}$ (Eq.~\eqref{hv_ng}), $ a_{\bg}$ and their fluctuations are indexed by $\bg$ (the set of N reciprocal lattice vectors) ordered in some fixed but arbitrary way. The $3$ dimensional vectors $\mathbf{j}$ (Eq.~\eqref{hv_j}) and $\bm{v}$ comprise of the three Cartesian components of the linear momentum density and their respective conjugates. Finally, $b$ is the thermodynamic conjugate to the internal energy density $e$ (Eq.~\eqref{hv_e}) of the system. Consequently, the static susceptibility matrix is a $(N+4)\times (N+4)$ dimensional matrix denoted by $\bm{\chi}$. The matrix $\bm{\chi}$ contains blocks representative of self and cross correlations between the different sets of hydrodynamic variables. 

\begin{subequations}\label{stat_cor0}
\begin{align}
\bm{\chi}(\bq)&=
\begin{bmatrix}
\bm{\chi}^{\rho\rho}_{(N\times N)} & \bm{\chi}^{\rho e}_{(N\times 1)} &\bm{\chi}^{\rho j}_{(N\times 3)} \\
\bm{\chi}^{e\rho}_{(1\times N)} & \bm{\chi}^{e e}_{(1\times 1)} &\bm{\chi}^{e j}_{(1\times 3)} \\
\bm{\chi}^{j\rho}_{(3\times N)} & \bm{\chi}^{je}_{(3\times 1)} & \bm{\chi}^{j j}_{(3\times 3)}
\end{bmatrix}\\
&=
\begin{bmatrix}
\bm{\chi}^{\rho\rho}_{(N\times N)} & \bm{\chi}^{\rho e}_{(N\times 1)} & 0 \\
\bm{\chi}^{e\rho}_{(1\times N)} & \bm{\chi}^{e e}_{(1\times 1)} & 0 \\
0 & 0 & \bm{\chi}^{j j}_{(3\times 3)}
\end{bmatrix}
\end{align}
\end{subequations}
The general form of $\bm{\chi}$ can be ascertained using arguments pertaining to the time reversal symmetries of density, energy and momentum~\cite{Forster}. While density and energy are even, momentum is odd with respect to time reversal. Quantities with opposite parity under time reversal cannot have non-zero static correlations {\it i.e.} the blocks $\bm{\chi}^{\rho j}$, $\bm{\chi}^{ej}$ and their complex conjugates are null matrices. Constituent components of the matrices $\bm{\chi}^{\rho\rho}_{(N\times N)}$, $\bm{\chi}^{\rho e}_{(N\times 1)}$ and $\bm{\chi}^{ee}_{(1\times 1)}$ are $\beta \langle \delta n^{*}_{\bg}(\bq)\delta n_{\bg'}(\bq)\rangle$, $\beta \langle \delta n^{*}_{\bg}(\bq)\delta e(\bq)\rangle$ and $\beta \langle \delta e^{*}(\bq)\delta e(\bq)\rangle$ respectively.

Acquiring explicit expressions of the variables $\delta a_{\bg}$, $\delta b$ and $\delta v_{\al}$ in terms of the relevant variables (see Eq.~\eqref{rel_conjugates}), requires $\bm{\chi}^{-1}$ and some shorthand definitions for the constituent matrix blocks of $\bm{\chi}$. The simple block diagonal form of the matrix (Eq.~\eqref{stat_cor0}) allows the independent inversion of the diagonal blocks (see Eq.~\eqref{Mat_id}a). The diagonal block $\bm{\chi}^{jj}_{(3\times 3)}$ involving the correlations between the momentum density fluctuations, can be inverted easily using the classical equipartition theorem $\langle p^{i}_{\alpha}p^{j}_{\beta}\rangle=mk_{B}T\delta_{ij}\delta_{\alpha\beta}$. With the mass per particle $m$ set to one, the correlations between the fluctuations of the densities of different components of linear momentum
\begin{align}\label{cl_eq_part}
   &\bm{\chi}^{j j}(\bq) =  \beta\langle\delta \hat{j}^{*}_{\al}(\bq)\delta \hat{j}_{\be}(\bq)\rangle =n_{0}V\delta _{\albe}
\end{align}
leads to the identification of the field conjugate to momentum; it is the velocity as given in Eq.~\eqref{rel_conjugates2}c (see Appendix~\ref{chi_inv}).
The block diagonal structure and the matrix identities in Eq.~\eqref{Mat_id} are used to perform the inversion 
%(see Appendix~\ref{chi_inv}) 
of the $(N\times N)$ submatrix in $\bm{\chi}$ and to derive the relation between the remaining pairs of thermodynamic conjugate variables 
\begin{subequations} \label{rel_conjugates2}
	\begin{align}
	&\daag(\bq,t) = \sum_{\bg'}J^{*}_{\bg\bg'}\delta n_{\bg'}(\bq,t) - U_\bg(\bq) \db(\bq,t)\\
	&\db(\bq,t)	= - L^{-1}(\bq) \sum_{\bg'} U^*_{\bg'}(\bq) \delta n_{\bg'}(\bq,t) + L^{-1}(\bq)\de(\bq,t)\\
    &\delta v_{\al}(\bq,t)= n^{-1}_{0}\delta j_{\alpha}(\bq,t). 
	\end{align}
\end{subequations}
Each of the terms $J_{\bg\bg'}$, $U_{\bg}$ and $L$ needs further interpretation. Given the general form (Eq.~\eqref{EOM_matrix1}a) for the components of the matrix $\bm{\chi}$, first we focus on the block $\bm{\chi}^{\rho\rho}_{(N\times N)}$. Let us define a $(N\times N)$ dimensional matrix ${\bf J}^{\rho\rho}$ such that ${\bf J}^{\rho\rho}=(\bm{\chi}^{\rho\rho})^{-1}$. This inverse density correlation matrix ${\bf J}^{\rho\rho}$ has components $J_{\bg\bg'}$ and is Hermitian. The Ornstein-Zernike relation (Eq.~\eqref{jgg}a) provides a connection between the density correlations and the inverse density correlation matrix. Moreover, the components $J_{\bg\bg'}$ can be obtained from the direct correlation function $c(\br_{1},\br_{2})$. Previous works~\cite{SG_jcp_2022,Haring2015} by the authors, explored this connection in great detail and here we present these relations for the sake of completeness.
\begin{subequations}\label{jgg}
\begin{align}
    k_{B}TV\delta_{\bg\bg''}=&\sum_{\bg'}\langle \delta \hat{\rho}^{*}_{\bg}(\bq,t)\delta \hat{\rho}_{\bg'}(\bq,t)\rangle J_{\bg'\bg''(\bq)}\\
    J_{\bg \bg^{'}}(\bq)=&\frac{k_{B}T}{V}\int d^{3}r_{1}\int d^{3}r_{2} e^{i\bg.\br_{1}}e^{-i\bg^{'}.\br_{2}}e^{i\bq.(\br_{1}-\br_{2})}\nonumber\\
    &\left[\frac{\delta(\br_{1}-\br_2)}{n(\br_{1})}-c(\br_{1},\br_{2})\right]
\end{align}
\end{subequations}
Next, we define $L$ and $U_{\bg}$ through the introduction of some short hand notations to represent components of specific correlations in the matrix $\bm{\chi}$ (Eq.~\eqref{stat_cor0})
\begin{subequations}\label{KLU_micro}
\begin{align}
    &K_{\bg}(\bq)=\beta\langle \delta \hat{e}^{*}(\bq)\delta \hat{\rho}_{\bg}(\bq)\rangle\\
    &K(\bq)=\beta\langle \delta \hat{e}^{*}(\bq)\delta \hat{e}(\bq)\rangle\\
    &L(\bq)=K(\bq)-\sum_{\bg\bg'}K_{\bg}(\bq)J_{\bg\bg'}(\bq)K^{*}_{\bg'}(\bq)\\
    &U_{\bg}(\bq)=\sum_{\bg'}J^{*}_{\bg\bg'}(\bq)K_{\bg'}(\bq)
\end{align}
\end{subequations}
The term $K_{\bg}(\bq)$ is the correlation between the fluctuations in energy density and Bragg peak amplitude around a reciprocal lattice vector $\bg$. $K(\bq)$ is the length ($\bq$) dependent second moment of the energy density. The inverse density correlation components $J_{\bg\bg'}(\bq)$ has been defined already in Eq.~\eqref{jgg}. Multiplicative combinations of these three terms show up as $L(\bq)$ (Eq.~\eqref{KLU_micro}c) and $U_{\bg}(\bq)$ (Eq.~\eqref{KLU_micro}d) in the components of the matrix $\bm{\chi}^{-1}$. In matrix representation, these constitute the matrix $\bm{\chi}^{-1}$ (see Appendix~\ref{chi_inv}) finally leading to the relations given in Eq.~\eqref{rel_conjugates2}. These microscopically defined quantities $L(\bq)$ and $U_{\bg}(\bq)$ will be revisited (see Eq.~\eqref{theta_tau_micro}), their small wave vector $q$ limits will be examined (see Eq.~\eqref{small_q_theta_tau}) and the implications of their symmetry properties will be discussed in section~\ref{coarse_grain_react}.

Incorporating the thermodynamic conjugates of $\hat{A}_{i}$ (see Eq.~\eqref{conj_MZ_1} and Eq.\eqref{rel_conjugates2}), the generalised Langevin equations Eq.~\eqref{EOM_gen1} can be re-written as 
\begin{align}\label{EOM_gen2}
     \langle\partial_{t}\hat{A}_{k}(t)\rangle^{\text{lr}}&=i\sum_{j}\omega^{*}_{kj}\langle\delta\tilde{A}_{j}(t)\rangle^{\text{lr}}\nonumber\\
     &-\sum_{j}\int_{0}^{t}d\tau m^{*}_{kj}(\tau-t)\langle\delta\tilde{A}_{j}(\tau)\rangle^{\text{lr}}.
\end{align}
Note the use of notations $\hat{A}_{i}$ and $\tilde{A_{i}}$ to denote microscopic dynamical variables and their respective thermodynamic conjugates.
%\com{NB: complex conjugation corresponds to $-\bq$ and thus is not required; actually, there should rather be a sum $\sum_{\bq}$. \\}
%\com{ After Eq 24 we should give the asymmetrical projector which averages with the extensive field to give the factor in front of the intensive one; (2.63) or (2.110) in Florian's thesis. \\}
%\onecolumngrid
\begin{widetext}
\begin{subequations}\label{project_micro}
  \begin{align}
     \mP
     &=\beta V^{-1}\sum_{\bq}\left(\sum_{\bg,\bg'}\delta n_{\bg}(\bq)\rangle J_{\bg\bg'}\langle\delta n^{*}_{\bg'}(\bq)
     +\delta e(\bq)\rangle L^{-1}(\bq) \langle\delta e^{*}(\bq)-\delta e(\bq)\rangle L^{-1}(\bq)\sum_{\bg''}U^{*}_{\bg''}(\bq)\langle \delta n^{*}_{\bg''}(\bq)+n^{-1}_{0}\delta j_{\al}(\bq)\rangle\langle \delta j^{*}_{\al}(\bq)\right)\\
     &=\beta V^{-1}\sum_{\bq}\left(\sum_{\bg'}\delta a_{\bg'}(\bq)\rangle\langle\delta n^{*}_{\bg'}(\bq)+\delta b(\bq)\rangle \langle\delta e^{*}(\bq)+\delta v_{\al}(\bq)\rangle\langle \delta j^{*}_{\al}(\bq)\right).
  \end{align}  
\end{subequations}
\end{widetext}
The projectors in Eq.~\eqref{project}, constructed from the set of relevant variables can now be given a more specific form for our system of interest. The Eq.~\eqref{project_micro} highlights the relation between the projection operator and the pairs of conjugate variables in Eq.~\eqref{rel_conjugates}.

Before we present Eq.~\eqref{EOM_gen2} for the crystalline solid with local defects, we simplify its terms further, in section~\ref{freq_micro}and section~\ref{mem_micro} to finally derive the equations of motion in section~\ref{TE_micro}.

\subsubsection{The frequency matrix}\label{freq_micro}
Once we have defined and simplified the components of the matrix $\bm{\chi}$, we attempt to first simplify and then evaluate the components of the $(N+4)\times (N+4)$ dimensional frequency matrix $\bm{\omega}$. This, as defined in Eq.~\eqref{EOM_matrix1}b, governs the reversible dynamical response of a system, pushed slightly out of equilibrium by an external perturbing field. Given the expression Eq.~\eqref{EOM_matrix1}b for the components of the frequency matrix, the symmetry of the relevant variables with respect to time-reversal  renders the components of $\bm{\omega}^{\rho\rho}_{(N\times N)}$, $\bm{\omega}^{\rho e}_{(N\times 1)}$, $\bm{\omega}^{ e e}_{(1\times 1)}$ and $\bm{\omega}^{j j}_{(3\times 3)}$ to zero \cite{Forster}. The rest of the components have been evaluated, from their microscopic definitions, in the Appendix~\ref{app_mat_cor} (Eq.~\eqref{omega_jrho} and Eq.~\eqref{omega_je}).
\begin{subequations}\label{mat_freq}
\begin{align}
    \bm{\omega}(\bq)&=
    \begin{bmatrix}
    \bm{\omega}^{\rho\rho}_{(N\times N)} &\bm{\omega}^{\rho e}_{(N\times 1)} &  \bm{\omega}^{\rho j}_{(N\times 3)}\\
    \bm{\omega}^{e\rho}_{(1\times N)} &\bm{\omega}^{ e e}_{(1\times 1)} &  \bm{\omega}^{ e j}_{(1\times 3)}\\
     \bm{\omega}^{j\rho}_{(3\times N)} &  \bm{\omega}^{je}_{(3\times 1)} &\bm{\omega}^{j j}_{(3\times 3)}
    \end{bmatrix}\\
    &=\begin{bmatrix}
    0 & 0 &  \bm{\omega}^{\rho j}_{(N\times 3)}\\
    0 & 0 &  \bm{\omega}^{ e j}_{(1\times 3)}\\
     \bm{\omega}^{j\rho}_{(3\times N)} &  \bm{\omega}^{je}_{(3\times 1)} & 0
    \end{bmatrix}
\end{align}
\end{subequations}
Since our calculations, seek to provide connections between microscopically derived equations of motion to well defined macroscopic thermodynamic variables, we need to interpret the different correlation functions in the small $\bq$ limit. To this end, we present here the definitions of the non-zero components $\bm{\omega}^{\rho j}_{(N\times 3)}$, $\bm{\omega}^{j\rho}_{(3\times N)}$, $\bm{\omega}^{ e j}_{(1\times 3)}$ and $\bm{\omega}^{je}_{(3\times 1)}$ of the frequency matrix at $\bq\rightarrow 0$
\begin{subequations}\label{freq_comp_micro}
\begin{align}
\bm{\omega}^{\rho j}_{\bg\al}(\bq)&=\beta\langle\delta \hat{\rho}^{*}_{\bg}(\bq)\mL\delta \hat{j}_{\al}(\bq)\rangle\nonumber\\
&=-V(g+q)_{\al}n^{*}_{\bg}+\mathcal{O}(q^{2})\\
\bm{\omega}^{j\rho}_{\al\bg}(\bq)&=\beta\langle\delta \hat{j}^{*}_{\al}(\bq)\mL\delta \hat{\rho}_{\bg}(\bq)\rangle\nonumber\\
&=-V(g+q)_{\al}n_{\bg}+\mathcal{O}(q^{2})\\
\bm{\omega}^{e j}_{\al}(\bq)&=\beta\langle\delta \hat{e}^{*}(\bq)\mL\delta \hat{j}_{\al}(\bq)\rangle\nonumber\\
&=-q_{\al}V \left( e_0 + p_0 \right)+\mathcal{O}(q^{2})\\
\bm{\omega}^{j e}_{\al}(\bq)&=\beta\langle\delta \hat{j}^{*}_{\al}(\bq)\mL\delta \hat{e}(\bq)\rangle\nonumber\\
&=-q_{\al}V \left( e_0 + p_0 \right)+\mathcal{O}(q^{2})
\end{align}
\end{subequations}
We evaluate them in terms of familiar thermodynamic parameters like energy and pressure. The explicit derivations of these terms from the microscopic expressions of the  correlations are given in the Appendix~\ref{app_mat_cor} (Eq.~\eqref{omega_jrho} and Eq.~\eqref{omega_je}). It is important to note that the expressions for $e_{0}$ and $p_{0}$ are achieved by taking the small wave vector $\bq$ limit of the microscopically defined, spatially varying energy and pressure, as depicted in the Appendix~\ref{app_e_p_eq}. The section~\ref{theory_coarse_graining} will once again take up the discussion of this frequency matrix $\bm{\omega}(\bq)$ after identifying, through coarse-graining, the connection between these definitions and the hydrodynamic definition of the analogous quantities in terms of the elastic fields like macroscopic density and displacements. 

\subsubsection{The memory matrix}\label{mem_micro}
Next we focus on the memory matrix $\bm{m}$ introduced in Eq.~\eqref{EOM_gen1} and defined in Eq.~\eqref{EOM_matrix1}c. Invoking the general continuity equation (Eq.~\eqref{cont_gen}) and the definition of the Liouville operator (Eq.~\eqref{def_L}), it becomes evident that the memory terms deal with the overlap between currents of the relevant variables. Therefore, these terms are expected to have finite correlation times. This again is a $(N+4)\times (N+4)$ dimensional Hermitian matrix separated into blocks similar to the susceptibility and the frequency matrix.
\begin{align}
    &\bm{m}(\bq,t)=
    \begin{bmatrix}
    {\bf m}^{\rho\rho}_{(N\times N)} & {\bf m}^{\rho e}_{(N\times 1)} &{\bf m}^{\rho j}_{(N\times 3)}\\
     {\bf m}^{*\rho e}_{(1\times N)} &{\bf m}^{ e e}_{(1\times 1)} &  {\bf m}^{ e j}_{(1\times 3)}\\
   {\bf m}^{*\rho j}_{(3\times N)} &  {\bf m}^{ * e j}_{(3\times 1)} &{\bf m}^{j j}_{(3\times 3)}
    \end{bmatrix}
\end{align}
The components of the constituent matrix blocks are given as follows 
\begin{align}\label{memory_terms_micro}
%    &m^{\rho\rho}_{\bg\bg'}(\bq,t)=\beta\langle \delta \hat{\rho}_{\bg}^{*}(\bq)\mL\mQ e^{-i\mQ\mL\mQ t}\mQ\mL\delta \hat{\rho}_{\bg'}(\bq) \rangle\\
    &m^{\rho e}_{\bg}(\bq,t)=\beta\langle \delta \hat{\rho}^{*}_{\bg}(\bq)\mL\mQ e^{-i\mQ\mL\mQ t}\mQ\mL\delta \hat{e}(\bq) \rangle.
%    &m^{\rho j}_{\bg\be}(\bq,t)=\beta\langle \delta \hat{\rho}_{\bg}^{*}(\bq)\mL\mQ e^{-i\mQ\mL\mQ t}\mQ\mL\delta \hat{j}_{\beta}(\bq) \rangle\\
%    &m^{ee}(\bq,t)=\beta\langle \delta \hat{e}^{*}(\bq)\mL\mQ e^{-i\mQ\mL\mQ t}\mQ\mL\delta \hat{e}(\bq) \rangle\\
%    &m^{ e j}_{\beta}(\bq,t)=\beta\langle \delta \hat{e}^{*}(\bq)\mL\mQ e^{-i\mQ\mL\mQ t}\mQ\mL\delta \hat{j}_{\beta}(\bq) \rangle\\ 
%    &m^{jj}_{\al\be}(\bq,t)=\beta\langle \delta \hat{j}^{*}_{\alpha}(\bq)\mL\mQ e^{-i\mQ\mL\mQ t}\mQ\mL\delta \hat{j}_{\beta}(\bq) \rangle
\end{align}
The explicit expression for $m^{\rho e}_{\bg}(\bq,t)$ is provided here and the rest of the memory terms $m^{\rho\rho}_{\bg\bg'}(\bq,t),m^{\rho j}_{\bg\be}(\bq,t),m^{ee}(\bq,t),m^{ e j}_{\beta}(\bq,t),m^{jj}_{\al\be}(\bq,t)$ are analogous in their expressions.
Since the memory terms represent the dissipative dynamics of the relevant variables and are interpreted as the overlaps between currents of conserved quantities, they are expected to have finite relaxation time scales. This we present as justification for our Markovian approximation~\cite{Zwanzig2001} to simplify our equations.
\begin{align}\label{markov_approx}
    \int_{0}^{t}d\tau m_{jk}(\bq,t-\tau)\Delta \tilde{A}_{j}(\tau)=\Gamma_{jk}(\bq)\Delta \tilde{A}_{j}(t)
\end{align}

In the small $\bq$ limit, the reduced dynamics of $\mQ\mL\mQ$ can be replaced (for proof see references~\cite{Forster,Flo_thesis}) with the dynamics of $\mL$. The Markovian approximation (Eq.~\eqref{markov_approx}) in the hydrodynamic limit $(\bq\rightarrow 0)$ gives the matrix comprising of Onsager transport coefficients presented in Eq.~\eqref{memory_micro_mkv}.
In the time evolution equation of the densities of the relevant variables, the reactive parts of the currents results from coupling (see Eq.~\eqref{EOM_matrix1}b) of the time derivative of one variable to another variable with opposite sign under time reversal. The dissipative parts, on the other hand, result from coupling of time derivatives of variables with same parity under time reversal. The Onsager transport coefficients $\Gamma_{jk}(\bq)$ follow from the Markovian approximation of the memory terms and are explicitly defined as follows
%\begin{subequations}
	\begin{align}\label{memory_micro_mkv}
%	\frac{\Gamma^{\rho\rho}_{\bg \bg'}(\bq)}{V} &=  \frac{\beta}{V} \int_0^\infty \dd t\, \scal{\delta \hat{\rho}_{\bg}^*(\bq) \liou \opq \ee^{-\ii \liou t} \opq \liou \delta \hat{\rho}_{\bg'}(\bq)} \\
	\frac{\Gamma^{\rho e}_{\bg}(\bq)}{V} &=   \frac{\beta}{V} \int_0^\infty \dd t\, \scal{\delta \hat{\rho}_{\bg}^*(\bq) \liou \opq \ee^{-\ii \liou t} \opq \liou \delta \hat{e}(\bq)}
%	\frac{\Gamma^{jj}_{\albe}(\bq)}{V} &=   \frac{\beta}{V} \int_0^\infty \dd t\, \scal{\delta \hat{j}_{\al}^*(\bq) \liou \opq \ee^{-\ii \liou t} \opq \liou \delta \hat{j}_{\be}(\bq)}\\
%	\frac{\Gamma^{ee}(\bq)}{V} &=   \frac{\beta}{V} \int_0^\infty \dd t\, \scal{\delta \hat{e}^*(\bq) \liou \opq \ee^{-\ii \liou t} \opq \liou \delta \hat{e}(\bq)}
	\end{align}
%\end{subequations}
Here we only give the example of $\Gamma^{\rho e}_{\bg}(\bq)$ as the other elements, $ \Gamma^{\rho \rho}_{\bg \bg'}(\bq),\Gamma^{jj}_{\albe}(\bq),\Gamma^{ee}(\bq)$ are built completely in analogy.
Physical significance of these transport coefficients will be examined in section~\ref{theory_coarse_graining} by seeking their relations to familiar quantities like viscosity, heat conductivity etc.

\subsubsection{Equations of motion}\label{TE_micro}
The equations of motion for the relevant variables, introduced in Section~\ref{theory_part1} Eq.~\eqref{rel_var_micro}, for the crystalline solid with a finite concentration of point-defects can now be obtained. Here, in addition to the simplifying approximations in the small wave vector limit, we invoke the thermodynamic conjugates, derived in Eq.~\eqref{rel_conjugates2}, of the relevant variables.
The relatively, simplified equations of motion for the microscopic relevant variables are given by 
%\com{please add $\Gamma^{j ..}$}
\begin{subequations}\label{EOM_micro_f}
	\begin{align}
	\partial_{t}\delta n_{\bg}(\bq,t) =& \left(\dfrac{\omega^{\rho j}_{\bg\al}(\bq)}{V} -\frac{\Gamma^{\rho j}_{\bg\al}(\bq)}{V}\right)\delta \va(\bq,t) \nonumber\\
	&- \sum_\bgp \frac{ \Gamma^{\rho\rho}_{\bg \bgp}(\bq)}{V}\dagp(\bq,t) -  \frac{\Gamma^{\rho e}_{\bg}(\bq)}{V} \db(\bq,t)\label{EOM_micro_fa}\\
	\partial_{t}\delta e(\bq,t) =& \left(\dfrac{\omega^{ej}_{\al}}{V}-\frac{\Gamma^{e j}_{\al}(\bq)}{V}\right) \delta\va(\bq,t) \nonumber\\
	&- \sum_\bg \frac{\Gamma^{e\rho}_{\bg}(\bq)}{V}\daag(\bq,t) -\dfrac{\Gamma^{ee}}{V}\db(\bq,t)\\
	\partial_{t}\delta \ja(\bq,t) =&-\dfrac{\Gamma^{jj}_{\al\be}}{V} \delta v_{\be}(\bq,t) \nonumber\\
	&+\left(\sum_{\bg}\dfrac{\omega^{j\rho}_{\al\bg}}{V}-\sum_{\bg}\frac{\Gamma^{ j\rho}_{\al\bg}(\bq)}{V}\right)\daag(\bq,t)\nonumber\\
&+\left(\dfrac{\omega^{je}_{\al}}{V}  -\frac{\Gamma^{je}_{\al}(\bq)}{V}\right)\delta b(\bq,t).
	\end{align}
\end{subequations}
These equations convey little intuition about the actual system without the frame of the system's thermodynamic properties to provide context. This connection is achieved by coarse-graining (see section~\ref{theory_coarse_graining}) the high dimensional space of the microscopic dynamics to the space of relevant thermodynamic variables. Besides substituting the components of the frequency matrix using Eq.~\eqref{freq_comp_micro}, we also show how the Onsager transport coefficients (Eq.~\eqref{memory_micro_mkv}) relate to constants (Eq.~\eqref{ons_coef_set1}) obtained from the dissipative dynamics in the system.

\section{The coarse-grained fields of elasticity}\label{micro_macro_fields}
\subsection{Coarse graining procedure}\label{theory_coarse_graining}
In order to derive connections between the microscopic equations of motions and the fields associated with the elasticity of a crystalline solid, the following ansatz was introduced in reference~\cite{Walz2010}.
\begin{align}\label{ansatz1}
\delta n_{\bg}(\bq,t) = -\ii \ng \ga \delta \ua(\bq,t) + \frac{\ng}{n_0} \delta n(\bq,t).
\end{align}
In an ideal crystal, the density fluctuation is equal to the divergence of the displacement field defined with respect to a fixed reference lattice. So the Bragg peak amplitude fluctuations of an ideal crystal at finite temperature would be given by the density fluctuations or the divergence of the displacement fields alone. This description breaks down in presence of mobile point defects like vacancies and interstitials. The ansatz in Eq.~\eqref{ansatz1} assumes that the fluctuations in the Bragg peak amplitudes in a crystal with local defects will have contributions from both divergence of displacements as well as density fluctuations. This description in the Fourier space also does not require any reference lattice for the definition of the displacement fields. 

Consistent with the ansatz Eq.~\eqref{ansatz1}, reference~\cite{Walz2010} proposed two linear combinations that lead to the standard thermodynamic variables, i.\,e. the number density and the displacement fields from the Bragg peak amplitudes. 
 \begin{subequations}\label{ansatz3_sum}
\begin{align}
\delta n(\bq,t) &= \frac{n_0}{\mathcal N_0} \sum_\bg \ng^* \delta n_{\bg}(\bq,t),\label{ansatz3_sum_a}  \\
\delta \ua(\bq,t) &= \ii \mathcal N^{-1}_\albe \sum_\bg \ng^* \gb \delta n_{\bg}(\bq,t), \label{ansatz3_sum_b}
\end{align}
\end{subequations}
The two normalizations in Eq.\eqref{ansatz3_sum} are $\mathcal N_0 = \sum_\bg \lvert \ng \rvert^2$ and $\mathcal N_\albe = \sum_\bg \lvert \ng \rvert^2 \ga \gb$. We get accordingly $\scal{n(\bq,t)} = N \delta_{\bq,0}$ and $\scal{\ua(\bq,t)}=0$. Also the symmetry argument, $\sum_g \lvert \ng \rvert^2 \gb=0$, essentially ensures orthogonality of $\delta n(\bq,t)$ and $\delta u_{\al}(\bq,t)$. Here we want to point out how our definition of the density field (Eq.~\eqref{ansatz3_sum_a}) is different from alternative approaches addressing the question of mechanical response in crystals with defects. The coarse-grained density fields in the theories developed by Haussmann~\cite{haussmann} and Mabillard \emph{et al.}~\cite{Mabi_1} arise from density fluctuations only around the centre of reciprocal space (viz.~at $\bg=0$) while we define the density field(Eq.~\ref{ansatz3_sum_a}) as a sum over fluctuations around all reciprocal lattice vectors in the system. Implications of this difference will become discernible only through future implementation of these approaches to study specific crystalline solids.

The large set of variables $\delta n_{\bg}(\bq)$ (Eq.~\eqref{hv_ng}), having been reduced to four coarse-grained variables $\delta n(\bq)$ and $\delta u_{\al=1,2,3}(\bq)$ through the ansatz Eq.~\eqref{ansatz1} leaves us with eight coarse-grained hydrodynamic variables. These variables, paired with their respective thermodynamic conjugates, give us a coarse-grained version of the Eq.~\eqref{rel_conjugates}
\begin{align}\label{rel_conjugates_cg}
    &\begin{bmatrix}
    \delta a(\bq,t)  \\
    \delta \mathbf{y}(\bq,t)\\
       \delta b(\bq,t)\\
    \delta \mathbf{v} (\bq,t) \\  
\end{bmatrix}
    =V\mathbf{\chi^{-1}}
\begin{bmatrix}
   \delta n(\bq,t)\\
   \delta \mathbf{u}(\bq,t)\\
      \delta e(\bq,t)\\
   \delta \mathbf{j} (\bq,t) \;
\end{bmatrix}.
\end{align}
Each pair of conjugate variables, along with reference to linear response relations, are derived and discussed in section~\ref{coarse_grain_react} as well as appendix~\ref{sec_chi_cg_app}. Attention is given to the vector $\delta \mathbf{u}$ constituted of $\delta u_{\al=1,2,3}$ in three dimensions since its relation to its conjugate $\delta \bm{y}$ with constituents $\delta y_{\al=1,2,3}$ leads to the coefficients of stiffness within linear elasticity.

\subsection{The coarse-grained projectors and coarse-grained intensive variables}\label{coarse_grain_react}
Now that we have established the ground work by deriving the important connections between the microscopic and the coarse-grained framework, we can go on to derive the equations of motion for the coarse-grained relevant variables (see Eq.~\eqref{EOM_cg}). The initial step, towards deriving and interpreting these equations of interest, is to obtain the coarse-grained projection operators which shall be denoted by $\tilde{\mP}$ and $\tilde{\mQ}$. For that, we need to revisit the thermodynamic conjugate variables in Eq.~\eqref{rel_conjugates} and Eq.~\eqref{rel_conjugates2}. Then define $\delta a$, $\delta b$ using the coarse-graining ansatz. Let us first establish the relations between the coarse-grained thermodynamic conjugates $\delta a$ and $\delta y_{\al}$ and their microscopic counterpart $\delta a_{\bg}$. Drawing analogy to Eq.~\eqref{ansatz3_sum}, these quantities are 
\begin{subequations}\label{cg_conj_def}
    \begin{align}
        &\delta a(\bq,t)=\dfrac{1}{n_{0}}\sum_{\bg}n^{*}_{\bg}\delta a_{\bg}(\bq,t)\\
        &\delta y_{\al}(\bq,t)=i\sum_{\bg}g_{\al}n^{*}_{\bg}\delta a_{\bg}(\bq,t).
    \end{align}
\end{subequations}
Next we substitute the $\delta n_{\bg}$ with the fluctuating fields $\delta n$ and $\delta u_{\al}$ using the ansatz in Eq.~\eqref{ansatz1} in the definitions of $\delta a_{\bg}$ and $\delta b$ given in Eq.~\eqref{rel_conjugates2} and repeated here with the substitutions,  
\begin{subequations} \label{rel_conjugates3}
	\begin{align}
	\daag(\bq,t) &= \dfrac{1}{n_{0}}\sum_{\bg'}J^{*}_{\bg\bg'}n_{\bg'}\delta n(\bq,t)-i\sum_{\bg'}J^{*}_{\bg\bg'}n_{\bg'}g'_{\al}\delta u_{\al}(\bq,t) \nonumber\\
	&- U_\bg(\bq) \db(\bq,t)\\
	\db(\bq,t)	&= - \dfrac{L^{-1}}{n_{0}} \sum_{\bg'} U^*_{\bg'}(\bq) n_{\bg'}\delta n(\bq,t) \nonumber\\
	&+iL^{-1} \sum_{\bg'} U^*_{\bg'}(\bq) n_{\bg'}g'_{\al}\delta u_{\al}(\bq,t)
	+ L^{-1}(\bq)\de(\bq,t).
	\end{align}
\end{subequations}
%\com{mf: changes in the following paragraphs}
The relations in Eq.~\eqref{rel_conjugates2} between the pairs of conjugate variables shown in Eq.~\eqref{rel_conjugates} requires the inversion of the static correlation matrix. Appendix~\ref{chi_inv}, Eq.~\eqref{stat_cor1}, explains how this involves the terms defined in Eq.~\eqref{KLU_micro}. Here, for deriving the relations given in Eq.~\eqref{conj_coarse-grained}, similar mathematical manipulations~\cite{Flo_thesis} are used in conjunction with the coarse-graining ansatz Eq.~\eqref{ansatz1}. 
 Plugging in the expression for $\delta a_{\bg}$ from Eq.~\eqref{rel_conjugates3}a to the equations in Eq.~\eqref{cg_conj_def} leads to expressions for $\delta a$ (Eq.~\eqref{conj_coarse-grained}a) and $\delta y_{\al}$ (Eq.~\eqref{conj_coarse-grained}c) in terms of the generalised elastic coefficients $\nu,\mu_{\al},\lambda_{\albe}$ given in Appendix~\ref{elastic_cons_app}. This allows us to write $\delta a, \delta b, \delta y_{\al}$ (Eq.~\eqref{conj_coarse-grained}) in terms of the fields of elasticity $\delta n, \delta u_{\al}$ and quantities like $\nu,\mu_{\al},\lambda_{\albe},\theta,\tau_{\al}$ which characterise the generalised material response of the solid. The relation between $\delta j_{\al}$ and its conjugate $\delta v_{\al}$ is repeated in this list to complete the set of relevant variables.
\begin{subequations}\label{conj_coarse-grained}
	\begin{align}
	\da(\bq,t)  
	=& \dfrac{\nu}{n^{2}_{0}}\delta n(\bq,t) -\dfrac{\mu_{\be}(\bq)}{n_{0}}\delta u_{\be}(\bq,t)\nonumber\\
	-&\frac{\theta^*(\bq)}{n_0} \db(\bq,t), \\
	\db(\bq,t) 
	=& L^{-1}(\bq) \left( -\frac{\theta(\bq)}{n_0} \dn(\bq,t) + \tau_\alpha(\bq) \dua(\bq,t)  \right) \nonumber\\
	+& L^{-1}(\bq)\de(\bq,t),\\ % \tag{\ref{eq:db'}} \\ %= \db(\bq,t)\\
 %\tag{\ref{eq:conj_var_energy_zm_mu}} \\
	\dya(\bq,t)  
	=& -\dfrac{\mu^{*}_{\al}(\bq)}{n_{0}}\delta n(\bq,t) +\lambda_{\al\be}(\bq)\delta u_{\be}(\bq,t)\nonumber\\
	+ &\tau^*_\alpha(\bq) \db(\bq,t)\\
	\delta v_{\al}(\bq,t)=& n^{-1}_{0}\delta j_{\alpha}(\bq,t)\label{conj_coarse-grained_d}
	\end{align}
\end{subequations}
%\com{It would be good to discuss the intensive variables more. Yet, the following sentence lacks evidence here.}
As already mentioned in section~\ref{theory_coarse_graining}, the variables $\delta y_{\al}(\bq)$ are components of the three dimensional vector ${\bf \delta y}$ which is conjugate to the three dimensional vector ${\bf \delta u}$ constituted of the displacement fields $\delta u_{\al}(\bq)$. The relation (Eq.~\eqref{ansatz3_sum}) of $\delta n$ and $\delta u_{\al}$ to the microscopic fields are reflected in their respective coarse-grained conjugates $\delta a$ and $\delta y_{\al}$. 
It would be pertinent here to note that the closed form representation (Eq.~\eqref{conj_coarse-grained}) of the coarse-grained conjugate fields $\delta a(\bq,t)$, $\delta b(\bq,t)$ and $\delta y_{\al}(\bq,t)$  require the use of a set of generalised elastic coefficients $\lambda_{\albe}(\bq), \mu_{\al}(\bq)$, and $\nu(\bq)$, which were derived and validated in previous papers. 
Appendix~\ref{mat_coef_app} summarises their microscopic definitions 
%of $\lambda_{\albe}(\bq), \mu_{\al}(\bq),\nu(\bq)$ 
for the sake of completeness.
References~\cite{SG_jcp_2022,Haring2015,Walz2010} present these important mechanical properties and by detailing their relation to the inverse density correlation matrix (Eq.~\eqref{jgg}), show ways of determining them in crystalline solids with local defects. Unlike this contribution, the previous works formulate the theory keeping only the reversible contributions (only the first term on the right hand side of Eq.~\eqref{EOM_gen1}) to the time evolution equations of the relevant variables and ignores all contributions of energy fluctuations or transport. They enter via two new coefficients
\begin{subequations}\label{theta_tau_micro}
\begin{align}
\tau_\alpha(\bq) &= \ii \sum_\bg U^*_\bg(\bq) \ng \ga \\
\theta(\bq) &= \sum_\bg U^*_\bg(\bq) \ng
\end{align}
\end{subequations}
defined using quantities given in Eq.~\eqref{KLU_micro}. Appendix ~\ref{co_tau_theta_L_app} discusses $\tau_{\al}(\bq)$, $\theta(\bq)$ to explain their microscopic origin and symmetries in the thermodynamic limit.
Given the definition of $\tau_{\al}(\bq)$ in Eq.~\eqref{theta_tau_micro}, Eq.~\eqref{rel_conjugates3}b transforms to Eq.~\eqref{conj_coarse-grained}b.

As material properties at long wavelengths are of interest to us, here we present the ${\bq}\rightarrow 0$ limit of the variables defined in Eq.~\eqref{KLU_micro} and Eq.~\eqref{theta_tau_micro}. The mathematical manipulations, starting with the initial definitions and leading to the final expressions, are detailed in Appendix~\ref{co_tau_theta_L_app}. Here, we recapitulate the final expressions for all the quantities $\tau_{\al}(\bq),\theta(\bq),L(\bq),\lambda_{\albe}(\bq),\mu_{\albe}(\bq),\nu(\bq)$ at $\bq\rightarrow 0$
\begin{subequations} \label{small_q_theta_tau}
	\begin{align}
	\lim_{\bq \rightarrow 0} \theta(\bq) &= \theta, \\
	\lim_{\bq \rightarrow 0} \tau_\alpha(\bq) &= \ii \qb \tau_\albe, \\
	\lim_{\bq \rightarrow 0} L(\bq) &= L.
	\end{align}
\end{subequations}
In the limit of macroscopic lengths, the leading order terms in $\theta(\bq)$ and $L(\bq)$ become $\bq$ independent real constants. $\tau_{\al}(\bq)$ at $\bq\rightarrow 0$, on the other hand, has symmetries similar to the generalised elastic constant $\mu_{\al}(\bq)$ which is associated~
\cite{Haring2015} with the coupling between the coarse-grained density $\delta n$ and displacement fields $\delta u_{\al}$ in a defect rich crystal; the symmetery holds $\tau_{\alpha\beta}=\tau_{\beta\alpha}$, as can be shown with the rotational LMBW equation~\cite{Haring2015,LMBW1,LMBW2}. As we show in the Appendix~\ref{elastic_cons_app}, the elastic coefficients $\lambda_{\albe}$, $\mu_{\al}$ and $\nu$ are the same as the ones in ~\cite{SG_jcp_2022,Haring2015,Walz2010} and in the hydrodynamic limit, they become
\begin{subequations}
\begin{align}
&\lim_{\bq \to 0}\lambda_{\alpha\beta}(\bq)=\lambda_{\alpha\beta\gamma\delta}q_{\gamma}q_{\delta}\\
&\lim_{\bq \to 0}\mu_{\alpha}(\bq)=i\mu_{\alpha\beta}q_{\beta}\\
&\lim_{\bq \to 0}\mu^{*}_{\alpha}(\bq)=-i\mu_{\alpha\beta}q_{\beta}\\
&\lim_{\bq \to 0}\nu(\bq)=\nu
\end{align}
\end{subequations}
Therefore, the conjugate variables in Eq.~\eqref{conj_coarse-grained}, in the $\bq\rightarrow 0$, are as follows 
\begin{subequations}\label{conj_coarse-grained_smallq}
	\begin{align}
	&\da(t) = \dfrac{\nu}{n^{2}_{0}}\delta n(t) -\dfrac{\mu_{\al\be}}{n_{0}} \delta u_{\alpha\beta}(t)-\frac{\theta}{n_0} \db(t) \\
	&\db(t) = L^{-1} \left( -\frac{\theta}{n_0} \dn(t) + \tau_\albe \delta u_{\albe}(t) + \de(t) \right)\label{conj_coarse-grained_smallq_b}\\ % \tag{\ref{eq:db'}} \\ %= \db(\bq,t)\\
 %\tag{\ref{eq:conj_var_energy_zm_mu}} \\
	&\delta y_{\albe}(t) =- \dfrac{\mu_{\al\be}}{n_{0}}\delta n(t)+\tau_\albe\db(t)\nonumber\\ &\ \ \ \ \ \ \ \ \ \ \ \ \ \ +\left(\lambda_{\al\gamma\be\delta}+\lambda_{\al\delta\gamma\be}-\lambda_{\al\be\gamma\delta}\right)\delta u_{\gamma\delta}(t)\label{conj_coarse-grained_smallqc}
	\end{align}
\end{subequations}
Here, 
\begin{align}\label{def_yab}
  \delta y_{\al} =-i\delta y_{\albe}q_{\be} 
\end{align}
has been used as the definition of $\delta y_{\albe}$ to obtain Eq.~\eqref{conj_coarse-grained_smallq}c from 
$\delta y_{\al}$ of Eq.~\eqref{conj_coarse-grained}c. Given the symmetries~\cite{Haring2015,SG_jcp_2022} $\mu_{\albe}=\mu_{\be\al}$, $\tau_{\albe}=\tau_{\be\al}$ and $\lambda_{\albe\gamma\delta}=\lambda_{\be\al\gamma\delta}=\lambda_{\albe\delta\gamma}=\lambda_{\gamma\delta\albe}$, the second rank tensor $\delta y_{\al\be}=\delta y_{\be\al}$ is also symmetric. The definition of $\delta y_{\al\be}$ in Eq.~\eqref{def_yab} is also a reference to the fact that in our calculations, we have chosen to restrict the displacement fluctuations to linear orders in strain by choosing $iq_{\beta}\delta u_{\al}(\bq\rightarrow 0)=\delta u_{\albe}$. Moreover, the orgin of the third term in Eq.~\eqref{conj_coarse-grained_smallqc} can be tracked~\cite{Flo_thesis,Haring2015} to the definition of the symmetric strain field (Eq.~\eqref{def_sym_uab}). Because of the symmetry of $\mu_{\albe},\tau_{\alpha\beta}$, as well as of $\lambda_{\alpha\beta\gamma\delta}$, only the symmetric part of the strain tensor enters.
%~\cite{SG_jcp_2022} the elastic free energy. This is a direct consequence of the independence of the linear elastic free energy on macroscopic torques on the system (see section~\ref{thermo_identity}). 
We therefore can redefine : 
\begin{align}\label{def_sym_uab}
 \delta u_{\albe} \to \frac 12 ( \delta u_{\albe}+ \delta u_{\beal})=\dfrac{1}{2}i(q_{\be}\delta u_{\al}+q_{\al}\delta u_{\be}).
 \end{align}
As conjugates to strains $\delta u_{\albe}$, the quantity $\delta y_{\albe}$ can be interpreted as linear stress variables. The relation between $\delta u_{\albe}$ and $\delta y_{\albe}$ as a pair of thermodynamic conjugates, is discussed in the context of thermodynamic free energy, in section~\ref{therm_conj_macro} (see Eq.~\eqref{rel_conjugates_cg} and Eq.~\eqref{conj_fs}).

Having derived the coarse-grained hydrodynamic variables and their thermodynamic conjugates, now we can define the coarse-grained projection operators that will allow us to obtain the macroscopic equilibrium and transport properties of the system.  The conservation law (see Eq.~\eqref{ceq_ng})
%\begin{align}\label{cons_ng}
%\partial_{t}\delta n_{\bg}(\bq)+i(g+q)_{\al}\delta j_{\al}(\bg+\bq)=0
%\end{align}
implies 
\begin{align}\label{cons_Qng}
    \mQ\mL\; \delta \hat\rho_{\bg}(\bq)=-(g+q)_{\al}\mQ\; \delta \hat j_{\al}(\bg+\bq)
\end{align}
where the microscopically defined projector $\mQ$ is acting on the mass conservation equation. Let us then take the time derivative of the coarse-graining ansatz in Eq.~\eqref{ansatz1} and look at the dynamics projected by $\mQ$
\begin{subequations}\label{ansatz2}
\begin{align}
&i\mQ\mL\delta n_{\bg}(\bq)=-i^{2}n_{\bg}g_{\al}\mQ\mL\delta u_{\al}(\bq)+\dfrac{n_{\bg}}{n_{0}}i\mQ\mL \delta n(\bq)  \\
\implies &\mQ \liou \dng(\bq) = - \ng \ga \tilde{\mQ} \ddua(\bq)
\end{align}
\end{subequations}
In Eq.~\eqref{ansatz2}a, the second term on the right, related to mass current $\mL\delta n$ vanishes if one postulates the conservation of number density $\partial_{t} \delta n=iq_{\al}\delta j_{\al}(\bq,t)$ and realises that the linear momentum densities belonging to the eigen-space of the projector do not contribute to the dynamics projected by $\tilde \mQ$. {\color{black}Arriving at Eq.~\eqref{ansatz2}b through this $\tilde \mQ\liou \dn(\bq)=-q_{\al}\tilde\mQ\delta j_{\al}(\bq) = 0$, uses the fact that the components of linear momentum $j_{\al}(\bq)$ 
%for $\bg=0$ 
in the 1st BZ are conserved variables. However, that does not automatically imply conservation of $j_{\al}(\bg+\bq)$ for $\bg\neq 0$.}

In Eq.~\eqref{ansatz2} we arrive at the relation between the microscopic projector $\mQ$ and $\tilde{\mQ}$ obtained after the coarse-graining ansatz Eq.~\eqref{ansatz1}. The coarse-grained space of a considerably smaller number of slow relevant variables, contains the number density $\delta n$, three components of linear momentum $\delta j_{\al}$, three components of displacement fields $\delta u_{\al}$ and energy density $\delta e$. These eight variables are compatible with the eight hydrodynamic modes expected~\cite{Fleming_Cohen} in an equilibrium crystalline solid at finite temperatures. This new smaller set of coarse-grained relevant variables require appropriate projectors written using them.  
\begin{widetext}
 \begin{align}\label{project_macro_cg}
     \tilde{\mP}=1-\tilde{\mQ}&=\beta V^{-1}\sum_{\bq}\left[\delta a(\bq)\rangle\langle\delta n^{*}(\bq)+\delta y_{\al}(\bq)\rangle\langle \delta u^{*}_{\al}(\bq)+\delta b(\bq)\rangle \langle\delta e^{*}(\bq)+\delta v_{\al}(\bq)\rangle\langle \delta j^{*}_{\al}(\bq)\right]
  \end{align} 
\end{widetext}
A comparison of this coarse-grained projector to the microscopic projector in Eq.~\eqref{project_micro} shows how all the terms involving fluctuations of Bragg peak amplitudes at the reciprocal lattice vectors, get included in the first two terms in the coarse-grained projector (Eq.~\eqref{project_macro_cg}) through the use of the ansatz Eq.~\eqref{ansatz1}. Before we can go on to obtain the dynamical equations of the coarse-grained variables, we need to determine the implication of the ansatz Eq.~\eqref{ansatz1} for the dissipative terms in the equations of motion (Eq.~\eqref{EOM_micro_f}).
%\com{again complex conjugation not required. this projector should be compared to (25).}

\subsection{The dissipative terms or the memory terms derived for the coarse-grained fields}\label{coarse_grain_dissi}

Equipped with the coarse-grained projectors $\tilde{\mP}$ and $\tilde{\mQ}$, we try to make sense of the transport coefficients in Eq.~\eqref{memory_micro_mkv}. We rewrite them with the coarse-grained projector and using the relation in Eq.~\eqref{ansatz2}b. % to derive the diffusion constants. Using the stress tensor, $\devt \ja = \ii \qb \sigab$, mass conservation, $\opq \liou \dn = 0$, and the energy current, $\devt \de = - \ii \qa \jea$, we can write
\begin{subequations}\label{ons_coef_set1}
	\begin{align}
	\frac{\Gamma^{*\rho\rho}_{\bg \bg'}(\bq)}{V} 
	&=  \frac{\beta}{V} \ng \ngp^* \ga \gb' \int_0^\infty \dd t\, \scal{\ddua^*(\bq) \tilde{\opq} \ee^{-\ii \liou t} \tilde{\opq} \ddub(\bq)}^* \nonumber\\
	&= \ng \ngp^* \ga \gb' \zeta_\albe, \\
	\frac{\Gamma^{*\rho e}_{\bg}(\bq)}{V} 
	&= \frac{\beta}{V} \ng \ga \qb \int_0^\infty \dd t\, \scal{\ddua^*(\bq) \tilde{\opq} \ee^{-\ii \liou t} \tilde{\opq} \jb^e(\bq)}^* \nonumber\\
	&= \ng \ga \qb \xi^\top_\albe, \\
	\frac{\Gamma^{*e\rho}_{\bg}(\bq)}{V} 
	&=  \frac{\beta}{V} \ng^* \gb \qa \int_0^\infty \dd t\, \scal{\ja^{e*}(\bq) \tilde{\opq} \ee^{-\ii \liou t} \tilde{\opq} \ddub(\bq)}^* \nonumber\\
	&= \ng^* \gb \qa \xi_\albe.
	\end{align}
\end{subequations}

Note how the $\mQ\mL\delta n_{\bg}$ replaced by $-n_{\bg}g_{\al}\tilde{\mQ}\delta \dot{u}_{\al}(\bq)$ in the Eq.~\eqref{ons_coef_set2} is related to the fluctuations of linear momentum at $\bg\neq 0$ through the conservation law in Eq.~\eqref{ceq_ng} and hence Eq.~\eqref{cons_Qng}. In the limit of long time and small wave vectors, the integrals in Eq.~\eqref{ons_coef_set1} represent the transport coefficients $\zeta_{\albe}$, $\xi_{\albe}$ and $\xi^{T}_{\albe}$. Their symmetries are consistent~\cite{Flo_thesis} with the Onsager reciprocal relations~\cite{Onsager_1,Onsager_2} dictated by the symmetry of the Hamiltonian under time-reversal. The physical significance of these transport coefficients will become evident when discussed, in section~\ref{sec_EOM_cg} and section~\ref{EOM_thermo}, in the context of  the dynamical equations of the hydrodynamic variables.

% This same quantity is represented by the time derivative of $\delta u_{\al}$ in Eq.~\eqref{ons_coef_set1}a and Eq.~\eqref{ons_coef_set1}b as a consequence of the coarse-graining ansatz in Eq.~\eqref{ansatz1}.
%{\color{red}what is the physical significance of this? The fact that our displacement field relates to the momentum density at finite $\bg$ appears to be a direct consequence of Eq.~\eqref{ansatz1}. what is its connection to the motion of point defects?}
Next consider the Onsager transport coefficient $\Gamma^{jj}_{\albe}(\bq)$ and then, in it, substitute the term $\mL\delta \hat{j}_{\al}(\bq)$ using the conservation of the linear momentum (Eq.~\eqref{ceq_j}). This leads to the viscosity tensor $\eta_{\albe\gade}$ given by the integral in the Eq.~\eqref{ons_coef_set2}. 
\begin{align}\label{ons_coef_set2}
\frac{\Gamma^{jj}_{\albe}(\bq)}{V}&=q_{\be}q_{\gamma}\eta_{\albe\gade}\nonumber\\
&=  \frac{\beta}{V} q_{\be}q_{\gamma} \int_0^\infty \dd t\, \scal{\sigab^*(\bq) \tilde\opq \ee^{- \ii \liou t} \tilde\opq \sigcd(\bq)}^*.
\end{align}
Another transport coefficient, arises from the memory term associated with the energy conservation (Eq.~\eqref{ceq_e})
\begin{align}\label{ons_coef_set3}
\frac{\Gamma^{ee}(\bq)}{V}&=q_{\al}q_{\be}\alpha_\albe T \nonumber\\
&=  \frac{\beta}{V}q_{\al}q_{\be} \int_0^\infty \dd t\, \scal{\ja^{e*}(\bq) \tilde\opq \ee^{-\ii \liou t} \tilde\opq \jb^e(\bq)}^*.
\end{align}\\
Note that both projectors $\opq$ and $\tilde\opq$ could be used  in Eqs.~\eqref{ons_coef_set2} and \eqref{ons_coef_set3} as they act identically in both kernels. The reason is time-parity.  

%\com{mf: I moved this}
The components of $\bm{\Gamma}^{\rho j}_{(N\times 3)}$, $\bm{\Gamma}^{ e j}_{(1\times 3)}$ and their conjugate transposes in the dynamical equations  Eq.~\eqref{EOM_micro_f} have been neglected in our calculations. It can be shown~\cite{Flo_thesis} that the leading $\bq$ dependent term for these components arises from $\mathcal{O}\left(\bq(\bg+\bq)\right)$ while all the other components of $\bm{\Gamma}$ has $\mathcal{O}(\bg+\bq)$ (see Eq.~\eqref{ons_coef_set1}, Eq.~\eqref{ons_coef_set2} and Eq.~\eqref{ons_coef_set3}) leading order terms.

After writing the complete equations of motions for the coarse-grained relevant variables in section~\ref{sec_EOM_cg}, physical significance of these transport coefficients will be discussed in section~\ref{sec_EOM_cg} as well as in section~\ref{EOM_thermo}.

%The equations of motion, therefore become
%\begin{subequations}
%	\begin{align}
%	\begin{split}
%	\partial_{t}\delta n_{\bg}(\bq,t) &= -\ii \ng \left( \ga + \qa \right) \va(\bq,t) \\
%	&\ghostequal- \zeta_\albe \sum_\bgp \ng \ngp^* \ga \gb'  \dagp(\bq,t)  \\
%	&\ghostequal - \qb \xi^\top_\albe  \ng \ga  \db(\bq,t)
%	\end{split} \\
%	\begin{split}
%	\partial_{t}\delta e(\bq,t) &= -\ii \left(e_0 +p_0 \right) \qa \va(\bq,t)  \\
%	&\ghostequal- \qa \xi_\albe \sum_\bg \ng^* \gb  \daag(\bq,t) \\
%	&\ghostequal - \qa \qb \alpha_\albe T \db(\bq,t) \end{split} \\
%	\begin{split}
%	\partial_{t}\delta \ja(\bq,t) &= -\ii \sum_\bg \ng^* \left( \ga + \qa \right) \daag(\bq,t)  \\
%	&\ghostequal- \ii \left(e_0 +p_0 \right) \qa \db(\bq,t)  \\
%	&\ghostequal- \qb \qc \eta_{\albe \gade} \vd(\bq,t) 
%	\end{split}
%	\end{align}
%\end{subequations}

\subsection{Equations of motion of the coarse-grained fields of elasticity}\label{sec_EOM_cg}
Incorporating all the coarse-grained variables in the microscopic equations of motion in Eq.~\eqref{EOM_micro_f} one obtains the full time evolution equations for the fields of elasticity. Appendix~\ref{eom_cg_app} explains the steps involved in the derivation of these final equations (Eq.~\eqref{EOM_cg}). This derivation of the hydrodynamic equations Eq.~\eqref{EOM_cg} in conjunction with the definitions of the material constants in section~\ref{coarse_grain_react}  and the Onsager transport coefficients in section~\ref{coarse_grain_dissi} provides a microscopic basis for all the reactive and dissipative coefficients that govern the macroscopic static and dynamic properties of crystalline solids with local-defects.
\begin{subequations}\label{EOM_cg}
    \begin{align}
    \partial_{t}\delta n(\bq,t) &= -\ii n_0 \qa \delta \va(\bq,t)\\
\partial_{t}\delta e(\bq,t) 
	&= -\ii \left(e_0 +p_0 \right) \qa \delta \va(\bq,t) \nonumber\\ 
	&- \qa \xi_\albe \delta y_{\be\gamma}(\bq,t) q_{\gamma} - \qa \qb \alpha_\albe T \db(\bq,t)\\
	\partial_{t}\delta \ja(\bq,t) &=-in_{0}q_{\al}\delta a(\bq,t)+i\delta y_{\al\be}(\bq,t)q_{\be}\nonumber\\
&- \ii \left(e_0 +p_0 \right) \qa \db(\bq,t) - \qb \qc \eta_{\albe \gade} \delta \vd(\bq,t)\\
\partial_{t}\delta {\ua(\bq,t)} &= \delta \va(\bq,t)\nonumber\\
&+i\zeta_\albe\delta y_{\be\gamma}(\bq,t)q_{\gamma} - \ii \qb \xi_\albe^\top \db(\bq,t).
    \end{align}
\end{subequations}

At this point, we have the time evolution equation of the eight slow variables in the system. Comparing the Eq.~\eqref{EOM_cg} to the analogous equations in case of a fluid~\cite{Forster,Kadanoff_Martin}, one immediately concludes that the terms with $\delta y_{\albe}$ would be absent in the fluid. In case of a fluid, there are only five hydrodynamic modes and a displacement field is ill-defined. So only five equations of motion specifying the time evolution of number density, energy density and linear-momentum density, would be relevant. This analogy to fluid equations will allow us to identify the viscosity $\eta_{\al\be}$ and transport coefficient for heat conductivity $\alpha_{\al\be}$.  We are left with two more dissipative constants $\zeta_{\albe}$ and $\xi_{\albe}$ %and $\xi^{T}_{\albe}$ 
each of which arises in an ordered solid where the displacement fields need to be treated as separate variables justified by the spontaneous breaking of translational invariance. This enters the possibility of defect motion into Eq.~\eqref{EOM_cg}, because the divergence of the displacement field is not tied to the density change as would hold in an ideal solid without defects.

\section{Connection to thermodynamics}\label{thermodynamics}

The original Mori-Zwanzig projection operator formalism, that eventually leads to these equations, does not provide a recipe for choosing an optimal set of slow variables that best captures the macroscopic properties of a given system. Conventionally, the variables associated with conservation laws or spontaneously broken continuous symmetries are found to be best suited for this purpose. Validating our chosen set of relevant variables, requires examination of how our equations of motion, in the small wave vector limit, relates to the thermodynamic parameters determining the equilibrium ensemble of the system. In section~\ref{thermodynamics}, we present a detailed derivation of the connections between the dynamics of the relevant variables and thermodynamics of the crystal with finite concentration of local defects.

\subsection{Free energy expansion and thermodynamic identities}\label{thermo_identity}
Any thermodynamic system at equilibrium, is characterised by a minimum in the free energy and a maximum in the entropy. Therefore, small changes in the thermodynamic parameters characterising the system, result in small increments in the free energy. Relying on this conceptual premise, the free energy density of a point-defect rich crystalline solid~\cite{chaikin_lubensky_1995}, can be expanded around the free energy density of a reference un-deformed $(%u_{\al\be}=
\nabla_{\be}u_{\al}=0)$ crystal phase at density and temperature $n_{0}$ and $T_{0}$ respectively. %\com{define the thermodynamic extensive variables, especially strain}
\begin{align}\label{F_nuT_exp}
&f (n_{0}+\delta n,0+\delta u_{\al\be},T_{0}+\delta T) \nonumber\\
=&f^0 + \mu^0 \dn + \frac{1}{2} A \left( \frac{\dn}{n_0} \right)^2 + B_\albe \delta u_{\albe} \frac{\dn}{n_0}\nonumber\\
+& \frac{1}{2} C^n_{\albe\gade} \delta u_{\albe}\delta u_{\gamma\delta}  - s^0 \dT + D_{s} \frac{\dn}{n} \dT \nonumber\\
+& \frac{1}{2} E \dT^2 + F_\albe \dT \cdot \delta u_{\albe},
\end{align}
The linear strain field $\delta u_{\al\be}=\dfrac{1}{2}(\nabla_{\be}u_{\al}+\nabla_{\al}u_{\be})$ is symmetrised in its definition because an asymmetry in the strain field in the thermodynamic limit amounts to a free energy conserving rotation of the entire system. The strain multiplied with the system volume $V$ is an extensive thermodynamic variable and we define $h_{\albe}$ as its conjugate intensive field. The coefficients of the quadratic couplings between the thermodynamic parameters are related to the curvature of the free energy along specific directions of the thermodynamic parameter space. Thus, the leading order terms in this expansion, provides important information related to the mechanical and thermal constants in the system.

The thermodynamic conjugates of the three independent variables $n$, $u_{\al\be}$ and $T$, are the chemical potential $\mu$, a tensor $h_{\al\be}$ and entropy density $s$.  Here, we are working with the free energy density $f=F/V$ and hence we use densities of the extensive variables like number of particles or entropy per volume. From the free energy expansion Eq.~\eqref{F_nuT_exp} and knowledge~\cite{chaikin_lubensky_1995} about the pair of thermodynamic conjugates, we get the following relations,
\begin{subequations}\label{conj_pheno}
	\begin{align}
	\dmu &= \frac{A}{n_0^2} \dn + \frac{B_\albe}{n_0}\delta u_{\albe} + \frac{D_{s}}{n_0} \dT \label{conj_pheno_a} \\
	\delta \hab &= \frac{B_\albe}{n_0} \dn + C^n_{\albe\gade}\delta u_{\gamma\delta} + F_\albe \dT  \label{conj_pheno_b}\\
	-\ds &= \frac{D_{s}}{n_0} \dn + F_\albe \delta u_{\albe} + E \dT. \label{conj_pheno_c} %\\
	%-n_0 \delta \bar s &= \frac{D}{n_0} \dn + F_\albe \uab + E \dT
	\end{align}
\end{subequations}
for the deviations of the conjugate fields around their equilibrium values, which are $\mu_0$ for the chemical potential, $s^0$ for the entropy density, and $h^0_{\albe}=0$ for the stress field conjugate to strain in this ensemble.
From these relations, one can  proceed to obtain the following thermodynamic derivatives and their Maxwell relations
\begin{subequations}\label{maxw_rel}
    \begin{align}
    &\dfrac{\partial^{2}f}{\partial n^{2}}\Big|_{u_{\albe},T}=\dfrac{\partial \mu}{\partial n}\Big|_{u_{\albe},T}=\dfrac{A}{n^{2}_{0}}\\
    &\dfrac{\partial^{2}f}{\partial n\partial u_{\alpha\beta}}=\dfrac{\partial \mu}{\partial u_{\albe}}\Big|_{n,T}=\dfrac{\partial h_{\albe}}{\partial n}\Big|_{u_{\gamma\delta},T}=\dfrac{B_{\albe}}{n_{0}}\label{maxw_rel_b}\\
    &\dfrac{\partial^{2}f}{\partial u_{\albe}\partial u_{\gamma\delta}}\Big|_{n,T}=\dfrac{\partial h_{\albe}}{\partial u_{\gamma\delta}}\Big|_{n,T}=C^{n}_{\alpha\beta\gamma\delta}\label{maxw_rel_c}\\
    &\dfrac{\partial^{2}f}{\partial n\partial T}=\dfrac{\partial \mu}{\partial T}\Big|_{u_{\albe},n}=-\dfrac{\partial s}{\partial n}\Big|_{u_{\gamma\delta},T}=\dfrac{D_{s}}{n_{0}}\label{maxw_rel_d}\\
    &\dfrac{\partial^{2}f}{\partial T^{2}}\Big|_{u_{\alpha\beta},n}=-\dfrac{\partial s}{\partial T}\Big|_{u_{\albe},n}=E\\
    &\dfrac{\partial^{2}f}{\partial T\partial u_{\alpha\beta}}=-\dfrac{\partial s}{\partial u_{\albe}}\Big|_{n,T}=\dfrac{\partial h_{\albe}}{\partial T}\Big|_{u_{\gamma\delta},n}=F_{\albe}\label{maxw_rel_f}
    \end{align}
\end{subequations}
The second derivatives of the free energy density provide coefficients of elasticity and coupling constants between different thermodynamic variables
which can be used in obtaining certain useful measurable quantities like isothermal compressibility, heat capacity per unit volume at constant volume and thermal expansion coefficient (see e.g.~\cite{Flo_thesis}).
One of the aims of this paper is to provide the microscopic basis to these material constants. % that show up in measurable quantities like isothermal compressibility or heat capacity etc. 
In the subsequent section~\ref{therm_conj_macro}, we present the connections between the constants that we identified in the relations Eq.~\eqref{maxw_rel} and the quantities that arise in the reactive and dissipative terms of the hydrodynamic equations we derived (Eq.~\eqref{EOM_cg}) using the projector Eq.~\eqref{project_macro_cg} after identifying the relevant dynamical variables and their respective thermodynamic conjugates(see Eq.~\eqref{rel_conjugates_cg}).

\subsection{Revisiting the Conjugate variables}\label{therm_conj_macro}
In section~\ref{theory}, when we first introduced the thermodynamic conjugate variables to the microscopic relevant variables in Eq.~\eqref{rel_conjugates} and then later re-framed them in Eq.~\eqref{conj_coarse-grained_smallq} using the coarse-graining ansatz, we never mentioned the physical significance of these quantities in the broader context of thermodynamic parameters. In this section and appendix~\ref{sec_chi_cg_app} we address this point.
%in a direct but seemingly ad hoc way \com{question?}. In the appendix~\ref{sec_chi_cg_app}, we use fluctuation-response arguments~\cite{chaikin_lubensky_1995,Forster,Haring2015} to verify the identifications.

In the appendix~\ref{sec_chi_cg_app}, we connect $(i)$ the change in the density of an extensive variable (like $n$) induced by the change in its intensive thermodynamic conjugate (like $\beta \mu$) (see Eq.~\eqref{app_chi_macro_2a}) to $(ii)$ correlations between the densities of the extensive thermodynamic variables (see Eq.~\eqref{app_chi_macro_2b}).  The matrix block ${\bm\chi_{nue}}$, corresponding to the correlations between fluctuations in density ($n$), displacement fields ${\bf u}$ with components $u_{\al}$ and energy density $e$, of the coarse-grained susceptibility matrix $\bm{\chi}$ is identified as matrix of thermodynamic derivatives; see  Eq.~\eqref{app_chi_macro_1}. These relations follow from evaluating the variances in the generalised grand-canonical ensemble  ~\cite{chaikin_lubensky_1995,Forster,Haring2015}. 
The final outcome of appendix~\ref{sec_chi_cg_app} is the identification of $\delta a$, $\delta {\bf y}$ and $\delta b$ in terms of the intensive thermodynamic quantities showing up in the expansion of the free energy in Eq.~\eqref{F_nuT_exp}. This we summarise here.
\begin{subequations}\label{conj_fs}
	\begin{align}
	&\da =\beta^{-1}\delta \left( \beta \mu \right)=T\delta \left( \frac{\mu}{T} \right) = \dmu - \frac{\mu^{0}}{T} \dT \\
	&\delta y_{\albe}=\beta^{-1}\delta (\beta h_{\albe})=T\delta \left(\dfrac{h_{\albe}}{T}\right)=\delta h_{\albe}, (\because h^{0}_{\albe}=0)\\
    &\db = -\beta^{-1}\delta \beta=T\delta\left(-\dfrac{1}{T}\right) = \frac{1}{T} \dT\label{conj_fs_c}.
	\end{align}
\end{subequations}
%In each case, the first equality gave the definition of the variables decorated with tildes.
%The coarse-grained projector in Eq.~\eqref{project_macro_cg}, which is written in terms of the relevant variables and their thermodynamic conjugates, can finally be fully interpreted with these %(Eq.~\eqref{conj_fs})  relations in the hydrodynamic limit. 
Using the information in Eq.~\eqref{conj_pheno_a} and Eq.~\eqref{conj_pheno_b}, we rewrite the expressions for $\delta a$,  $\delta y_{\albe}$ and $\db$
\begin{subequations}\label{conj_pheno_compare}
\begin{align}
	&\delta a=\delta \mu-\dfrac{\mu_{0}}{T}\delta T = \dfrac{A}{n^{2}_{0}}\delta n +\dfrac{B_{\al\be}}{n_{0}} \delta u_{\alpha\beta}+\frac{D_{s}}{n_0}\delta T-\mu_{0}\dfrac{\delta T}{T}, \\
	&\delta y_{\albe}=\delta h_{\albe}  = \dfrac{B_{\al\be}}{n_{0}}\delta n+C^{n}_{\albe\gamma\delta}\delta u_{\gamma\delta}+F_{\albe}\dT.
\end{align}
\end{subequations}
and then compare them to Eq.~\eqref{conj_coarse-grained_smallq}
to obtain 
\begin{subequations}\label{compare_pheno_micro}
\begin{align}
& \dfrac{\partial \mu}{\partial n}\Big|_{u_{\albe},T}=\dfrac{A}{n^{2}_{0}}=\dfrac{\nu}{n^{2}_{0}}\\
& \dfrac{\partial \mu}{\partial u_{\albe}}\Big|_{n,T}=\dfrac{B_{\albe}}{n_{0}}=-\dfrac{\mu_{\albe}}{n_{0}}\label{compare_pheno_micro_b}\\
& \dfrac{\partial h_{\albe}}{\partial u_{\gamma\delta}}\Big|_{n,T}=C^{n}_{\albe}=\left(\lambda_{\al\gamma\be\delta}+\lambda_{\al\delta\gamma\be}-\lambda_{\al\be\gamma\delta}\right)
\end{align}
\end{subequations}
Here we have invoked the set of Eq.~\eqref{conj_pheno} and Eq.~\eqref{maxw_rel} for the relations in Eq.~\eqref{compare_pheno_micro}. First we focus on the coefficients of $\delta n$ and $\delta u_{\albe}$ and take up the identification of the coefficients to $\delta T$ separately. 
%In order to ensure consistency, we present  the identification of $D_{s},E,F_{\albe}$ only after deriving an expression for the entropy density $\delta s$ in terms of $\theta,L,\tau_{\albe}$ (see Eq.~\eqref{theta_tau_L_thermo}). Comparison of this equation to Eq.~\eqref{conj_pheno_c} allows us to obtain the relations in Eq.~\eqref{theta_tau_L_Tf}.
The thermodynamic relation given in Eq.~\eqref{maxw_rel_b} shows $B_{\albe}$ to be a coupling between the displacement and the density fields. Previous theoretical perspectives derived by Szamel \emph{et al.}~\cite{Szamel_Ernst_1993,Szamel_1997} neglect the contribution of this coupling to the linear elastic response of crystals with point-defects. However, following the definition given by Walz \emph{et al.}~\cite{Walz2010}, later studies~\cite{Haring2015,lin2021,SG_jcp_2022} evaluate this quantity in isothermal crystalline systems~\cite{Mladek2006,Mladek2007,Likos2007,lin2021} with known direct correlation functions from classical density functional theory. In this paper, with a more general treatment of the thermodynamics, we recover (see Eq.~\eqref{compare_pheno_micro_b}) the definition of this coupling term while identifying similar cross-correlations of density and displacements with temperature (see Eq.~\eqref{maxw_rel_d} and Eq.~\eqref{maxw_rel_f}). We explain (see Eq.~\eqref{theta_tau_L_Tf}) how these quantities can be defined in terms of coefficients derived from microscopic fluctuations.

At this stage, before proceeding to re-consider the interpretations of the equations of motion (Eq.~\eqref{EOM_cg}), we need thermodynamic basis for the terms $\theta,\tau_{\albe}$ and $L$, that we introduced in our derivations of the dynamical equations. So we examine the connections between $\theta,\tau_{\albe},L$ and $D_{s},E,F_{\albe}$ (Eq.~\eqref{maxw_rel}). Following from the first and the second laws of thermodynamics, the relation between the density of entropy and internal energy for a system at constant volume and no external strain, is given by~\cite{Forster}
\begin{align}\label{rel_s_e_n}
\frac{1}{T} \de = \ds + \frac{\mu}{T} \dn.
\end{align}
This relation is identical to the ones employed in the context of the hydrodynamic description of a simple one-component fluid~\cite{Forster}.
Now, let us reconsider the variable $\delta b$ (Eq.~\eqref{conj_coarse-grained_smallq_b}), which now has been established (Eq.~\eqref{conj_fs_c}) as an intensive thermodynamic field. In Eq.~\eqref{conj_coarse-grained_smallq_b}, substitute $\delta b$ and $\delta e$ using Eq.~\eqref{conj_fs_c} and Eq.~\eqref{rel_s_e_n} respectively. Rearranging the substituted equation gives Eq.~\eqref{theta_tau_L_thermo}. Compare this expression for $\delta s$ with Eq.~\eqref{conj_pheno_c} to obtain relations between microscopically defined variables like $\theta$, $\tau_{\al\be}$, $L$ and the thermodynamic coefficients in the expansion of the free energy density (Eq.~\eqref{F_nuT_exp}). 
%\begin{subequations}\label{theta_tau_L_thermo}
\begin{align}
%&\dfrac{\delta T}{T} = L^{-1} \left( -\frac{\theta}{n_0} \dn + \tau_\albe \delta u_{\al\be} + \de \right)\\
-&\delta s =\left(\dfrac{\mu}{T}-\dfrac{\theta}{n_{0}T}\right)\delta n -\dfrac{L}{T^{2}}\delta T+\dfrac{\tau_{\albe}}{T}\delta u_{\albe}\label{theta_tau_L_thermo}
\end{align}
%\end{subequations}
Equating the coefficients of $\delta n, \delta T$ and $\delta u_{\albe}$ in the equations Eq.~\eqref{theta_tau_L_thermo} and Eq.~\eqref{conj_pheno_c} leads to the following relations
\begin{subequations}\label{theta_tau_L_Tf}
	\begin{align}
	&\theta = \mu n_0 - T D_{s}  \\ % = e+p -TD, \\
	& L = -T^2 E\\
	&\tau_\albe = TF_\albe. % \mathred{-\hab}, \\.
	\end{align}
\end{subequations}
%\com{mf: is the next paragraph useful or should we delete until 'This concludes..'??}
%This section~\ref{therm_conj_macro} provides the the thermodynamic context to all the variables that arise in our microscopic theory (section~\ref{theory}). Besides providing the relations in Eq.~\eqref{conj_fs}, here the thermodynamic basis for the quantities $\nu,\mu_{\albe},\lambda_{\albe\gamma\delta},\theta, \tau_{\albe},L$ (see Eq.~\eqref{compare_pheno_micro} and Eq.~\eqref{theta_tau_L_Tf}) are obtained using the identities derived in section~\ref{thermo_identity} (see Eq.~\eqref{conj_pheno} and Eq.~\eqref{maxw_rel}). Here we also present a summary of the results obtained in appendix~\ref{sec_chi_cg_app} where we verify the identifications of the elements of the inverse susceptibility matrix as thermodynamic derivatives. The matrix $\chi$ itself is related to thermodynamic derivatives, so that its inverse $\chi^{-1}$ can be interpreted by changing the ensemble. 
This concludes the thermodynamic interpretation of all the microscopically defined variables derived and used in the equations of motion in section~\ref{sec_EOM_cg}.
%Or if we prefer, the temperature and the chemical potential are expressed by
%\begin{subequations}
%\begin{align}
%\dT(\bq,t) &= T \db(\bq,t)\\
%\dmu(\bq,t) &= \da(\bq,t) + \mu \db(\bq,t)
%\end{align}
%\end{subequations}

%\com{Our ansatz 33 enters the coarse-grained susceptibility 53b. I wonder whether the difference in the coarse-grained density field to Rudolf's can be discussed with $\chi$ and $\chi^{-1}$?}

\subsection{Equations of motion}\label{EOM_thermo}
In this section we want to re-examine the hydrodynamic equations (Eq.~\eqref{EOM_cg}) with the knowledge of the thermodynamic relations derived in section~\ref{therm_conj_macro}. Inserting the conjugate variables $\delta a$, $\delta b$ and $\delta y_{\al}$ from Eq.~\eqref{conj_fs} into the time evolution equations (Eq.~\eqref{EOM_cg}) of the relevant variables, one obtains
\begin{subequations}\label{EOM_olt}
\begin{align}
\partial_{t}\delta n(\bq,t)&=-in_{0}q_{\al}\delta v_{\al}(\bq,t)\\
  \partial_{t}\delta e(\bq,t) &= -\ii \left(e_0 +p_0 \right) \qa \delta v_{\al}(\bq,t) \nonumber\\
  &+ q_{\al}q_{\gamma}\xi_\albe \delta h_{\gamma\be}(\bq,t)- \qa \qb \alpha_\albe \delta T (\bq,t)\\
 \partial_{t}{\delta \ja(\bq,t)} &= \ii \qb \delta h_{\al\be}(\bq,t)-\ii \qb\delta p(\bq,t)\delta_{\al\be} \nonumber\\
 &- \qb \qc \eta_{\albe\gade}\delta v_{\delta}(\bq,t)\label{EOM_oltc}\\
 \partial_{t}{\delta \ua(\bq,t)} &=\delta v_{\al}(\bq,t) \nonumber\\
& +iq_{\gamma}\zeta_\albe \delta h_{\gamma\beta}(\bq,t) - \ii \qb \xi_\albe^\top \dfrac{\delta T(\bq,t)}{T}\label{EOM_oltd}
\end{align}
\end{subequations}
Acquiring this version of the equation for the density of the linear momentum (Eq.~\ref{EOM_oltd}) required further simplification. First, using $p= -e + \mu n + Ts$ and the Gibbs--Duhem relation, $\ddp = n\dmu + s\dT$, one arrives at 
%Following this reasoning, we can also assume that $\dca = \dva - \frac{\va^0}{T} \dT = \dva$.
\begin{subequations}
	\begin{align}
	n_0 \da + (e_0 + p_0) \db &= n_0 \dmu + \frac{e_0+p_0-\mu n_0}{T} \dT \\
	&=\ddp.
	\end{align}
\end{subequations}
which simplifies the equation of motion of the momentum density given in Eq.~\eqref{EOM_cg}c to the following
\begin{align}
\partial_{t}{\delta \ja(\bq,t)} =& -\ii \qa \delta p(\bq,t)-\delta y_{\al}(\bq,t) \nonumber\\
&- \qb \qc \eta_{\albe\gade} \vd(\bq,t).
\end{align}

As we additionally identified all the intensive thermodynamic variables conjugate to their extensive counterparts (see Eqs.~\eqref{conj_fs}),
the final hydrodynamic equations of motion of a real crystal have been obtained starting from microscopic variables. Now,
we can consider their reactive and dissipative contributions. One of the obvious outcomes of the reactive couplings between momentum density and number density as well as the displacement fields, are the coefficients of elasticity. These coefficients, derived from the microscopic perspective of Mori-Zwanzig formalism has been discussed in great detail in references~\cite{SG_jcp_2022, Walz2010, Haring2015}. The focus of these previous studies were the derivation of the isothermal reversible reactive couplings between the relevant hydrodynamic variables and their proper validation through implementation in appropriate local-defect rich crystalline solids. In this contribution, in addition to deriving the microscopic basis for the dissipative terms, we also obtain the equation of motion of the energy density (Eq.~\eqref{EOM_olt}b).

\subsection{Implications of our results}
The phenomenon of heat transport in a solid medium at finite temperature is observed to be diffusive. Consistent with existing macroscopic theory perspectives~\cite{Fleming_Cohen,Mabi_1}, our equations yield diffusive heat transport with two separate transport coefficients $\xi_{\al\be},\alpha_{\albe}$ (see Eq.~\eqref{heat_transport}c) and their respective microscopic definitions (Eq.~\eqref{ons_coef_set1}c and Eq.~\eqref{ons_coef_set3}). For an equilibrated fluid or a crystalline solid with constant point-defect concentration and in absence of any external deforming fields, one can define a variable $\hat{q}(\br,t)=\hat{e}(\br,t)-\dfrac{e_{0}+p_{0}}{n_{0}}\hat{n}(\br,t)$ whose fluctuation, in the thermodynamic limit, is
\begin{align}
    \delta q=\delta e-\dfrac{e_{0}+p_{0}}{n_{0}}\delta n.
\end{align}
It can be shown~\cite{Forster,Kadanoff_Martin} that $\delta q$ is related to the entropy $S$ of the system through the relation $\delta q=Tn_{0}\delta (S/N)$. Thus, following the second law of thermodynamics, $q$ can be interpreted as the heat density in the system. Rearranging the terms in the Eq.~\eqref{EOM_olt}b for the energy density
\begin{widetext}
\begin{subequations}\label{heat_transport}
\begin{align}
&\partial_{t}\delta e(\bq,t) +\ii \left(e_0 +p_0 \right) \qa \delta v_{\al}(\bq,t) =q_{\al}q_{\be}\xi_\albe \delta h_{\al\be}(\bq,t)- \qa \qb \alpha_\albe \delta T (\bq,t)\\
\implies&\partial_{t}\delta e(\bq,t) - \dfrac{e_0 +p_0}{n_{0}} \partial_{t}\delta n(\bq,t) =q_{\al}q_{\be}\xi_\albe \delta h_{\al\be}(\bq,t)- \qa \qb \alpha_\albe \delta T (\bq,t)\\
\implies&\partial_{t}\delta q=q_{\al}q_{\be}\xi_\albe \delta h_{\al\be}(\bq,t)- \qa \qb \alpha_\albe \delta T (\bq,t)
\end{align}
\end{subequations}
\end{widetext}
%and substituting 
where $\delta v_{\al}(\bq)$ was substituted using the mass conservation equation Eq.~\eqref{EOM_olt}a, one ends up with the two dissipative contributions (Eq.~\eqref{heat_transport})c to the diffusion of heat in a crystal with point defects. Here we point out the distinctive aspects of the dissipative terms which distinguish a crystalline solid from a fluid. The analogue to the dissipative coefficient tensor $\alpha_{\albe}$ in a fluid is a scalar representing heat conductivity. In case of the crystalline solid $\alpha_{\albe}$ is a second rank tensor whose symmetry is determined by the symmetry of the concerned crystalline structure. Moreover, the additional dissipative coupling term $\xi_{\al\be}$, arising due to the coupling between energy and displacement fields, appears in the time evolution equation for energy (Eq.~\eqref{EOM_olt}b). Because of Onsager symmetry, its transpose $\xi_{\al\be}^{T}$, appears in the time evolution equation for the displacement fields Eq.~\eqref{EOM_olt}d. The tensor $\xi_{\al\be}$ %and $\xi_{\al\be}^{T}$ are 
is characteristic for a system with long range order. 

In a crystalline solid, the two diffusive hydrodynamic modes are heat transport and diffusion of point defects~\cite{Martin_Parodi_Pershan,Fleming_Cohen}. Remaining consistent with our previous contributions~\cite{Walz2010,Haring2015,SG_jcp_2022}, we define the field of fluctuating point-defect concentration as 
\begin{align}\label{defect_conc}
    \delta c(\bq,t)=-\delta n(\bq,t)-n_{0}\ii q_{\al}\delta u_{\al}(\bq,t).
\end{align}
Taking the time derivative of Eq.~\eqref{defect_conc}, then substituting $\partial_{t}\delta n(\bq,t)$ and $\partial_{t}\delta u_{\al}(\bq,t)$ using Eq.~\eqref{EOM_olt}a and Eq.~\eqref{EOM_olt}d immediately shows that the ansatz Eq.~\eqref{defect_conc} leads to the following 
\begin{align}\label{defect_diff}
    \partial_{t}\delta c(\bq,t)=n_{0}q_{\al}q_{\gamma}\zeta_\albe \delta h_{\gamma\beta}(\bq,t) - n_{0}q_{\al} \qb \xi_\albe^\top \dfrac{\delta T(\bq,t)}{T}
\end{align}
equation for the time evolution of the point-defect concentration. %We choose not to derive our hydrodynamic equations using $\delta q(\bq,t)$ and $\delta c(\bq,t)$ as relevant variable alternatives to $\delta e$ and $\delta n$ respectively. However, 
Eq.~\eqref{heat_transport}c and Eq.~\eqref{defect_diff} show that the leading order contributions to the time evolution equations for the heat density and the point-defect density are of the order $\bq^{2}$. This is consistent with the phenomenological theories~\cite{Fleming_Cohen,Mabi_1} and predicts the transport processes for heat and point-defects in a crystalline solid to be diffusive. 

The definition of the fluctuation of the point-defect concentration (Eq.~\eqref{defect_conc}) will help to rationalize how the present, more general, perspective reduces to our previous microscopic dissipation-less description of isothermal elastic properties of a point-defect rich crystal \cite{Haring2015}. Substitution of the density fluctuation $\delta n(\bq,t)$ in terms of the defect density concentration $\delta c(\bq,t)$ and the displacement fields $\delta u_{\al}(\bq,t)$ in Eq.~\eqref{EOM_micro_macro_j_app}a, followed by some mathematical manipulations~\cite{Flo_thesis}, results in the following equation for the density of linear momentum 
\begin{align}\label{EOM_mom_def}
  \partial_{t}\delta j_{\al}(\bq,t)=&-\Lambda_{\albe}(\bq)\delta u_{\beta}(\bq,t)\nonumber\\
  &-V_{\al}(\bq)\delta c(\bq,t)-\dfrac{Z^{*}_{\al}(\bq)}{T}\delta T (\bq,t) 
\end{align}
where % \com{please only give the leading orders with $\mu_\albe$ etc.}
\begin{subequations}\label{lam_V_Z}
\begin{align}
\Lambda_{a\be}(\bq)&=\lambda_{\albe}(\bq)-iq_{\al}\mu_{\be}(\bq)+iq_{\be}\mu^{*}_{\al}(\bq)+q_{\al}\nu(\bq)q_{\be}\nonumber\\
&\approx\Lambda_{\albe\gamma\delta}q_{\beta}q_{\delta}\label{lam_V_Z_a}\\
\Lambda_{\albe\gamma\delta}&=\lambda_{\al\gamma\be\delta}+\lambda_{\al\delta\gamma\be}-\lambda_{\al\be\gamma\delta}\nonumber\\
&+\delta_{\al\be}\mu_{\gamma\delta}+\mu_{\albe}\delta_{\gamma\delta}+\nu \delta_{\albe}\delta_{\gamma\delta}\\
V_{\al}(\bq)&=\dfrac{1}{n_{0}}\left[\mu^{*}_{\al}(\bq)-iq_{\al}\nu(\bq)\right]\nonumber\\
&\approx \dfrac{1}{n_{0}}\left[-i\mu_{\albe}q_{\be}-i\nu q_{\al}\right]\label{lam_V_Z_c}\\
Z^{*}_{\al}(\bq)&=\tau^{*}_{\al}(\bq)+iq_{\al}\left[e_{0}+p_{0}-\theta^{*}(\bq)\right]\nonumber\\
&\approx -i\tau_{\albe}q_{\be}+iq_{\al}[e_{0}+p_{0}-\theta].\label{lam_V_Z_d}
\end{align}
\end{subequations}
The second lines in the expressions for $\Lambda_{\albe}(\bq)$ (Eq.~\eqref{lam_V_Z_a}), $V_{\al}(\bq)$ (Eq.~\eqref{lam_V_Z_c}) and $Z^{*}_{\al}(\bq)$ (Eq.~\eqref{lam_V_Z_d}) represent the leading terms in the $\bq\rightarrow 0$ limit of the respective quantities. Since the aim is to show how to recover our previous results from a more general description, we have chosen to keep only the reversible parts in Eq.~\eqref{EOM_mom_def}. In this limit, Eq.~\eqref{EOM_oltd} reduces to $\partial_t \delta \bu =\bv$ in which $\bv$ is related to ${\bf j}$ through the relation in Eq.~\eqref{conj_coarse-grained_d}. If the system is isothermal, setting $\delta T$ to zero, tive-derivative of Eq.~\eqref{EOM_mom_def} leads to the isothermal wave equation for the  linear momentum, first derived in reference~\cite{Walz2010}. Moreover, the quantity $\Lambda_{\albe}(\bq)$ is identical to the dynamical matrix associated with the isothermal elastic properties of the crystal with constant defect concentration. Thus the definitions of the quantities $\Lambda_{\albe}(\bq)$ and $V_{\al}(\bq)$ given in Eq.~\eqref{lam_V_Z} have the same interpretations as in references~\cite{Walz2010,Haring2015,SG_jcp_2022}. 

If one chooses to write the reversible part of the density of the linear momentum while considering $\delta q(\bq,t)$ and $\delta c(\bq,t)$ as the relevant variables, the following equation is obtained \begin{align}\label{EOM_j_adia}
  \partial_{t}\delta j_{\al}(\bq,t)&=-\Lambda^{ad}_{\albe}(\bq)\delta u_{\beta}(\bq,t)\nonumber\\
  &-V^{ad}_{\al}(\bq)\delta c(\bq,t)-L^{-1}Z^{*}_{\al}(\bq) \delta q    
\end{align}
where %\com{please also give the $q\to0$ limits, as described in the sentence below (70).}
\begin{subequations}\label{lam_V}
\begin{align}
&\Lambda^{ad}_{a\be}(\bq)\nonumber\\
&=\Lambda_{\albe}(\bq)+L^{-1}(\bq)Z^{*}_{\al}(\bq)Z_{\be}(\bq)\nonumber\\
&\approx \Lambda_{\albe\gamma\delta}q_{\be}q_{\delta}\nonumber\\
&+L^{-1}(\tau_{\al\gamma}q_{\gamma}-q_{\al}[e_{0}+p_{0}-\theta])(\tau_{\be\delta}q_{\delta}-q_{\be}[e_{0}+p_{0}-\theta])\\
&V^{ad}_{\al}(\bq)\nonumber\\
&=V_{\al}(\bq)-L^{-1}(\bq)Z^{*}_{\al}(\bq)\left(\dfrac{e_{0}+p_{0}-\theta^{*}(\bq)}{n_{0}}\right)\nonumber\\
&\approx \dfrac{-i}{n_{0}}(\mu_{\albe}q_{\be}-\nu q_{\al})\nonumber\\
&+\dfrac{i}{n_{0}}L^{-1}\left(\tau_{\albe} q_{\be}-q_{\al}[e_{0}+p_{0}-\theta]\right)(e_{0}+p_{0}-\theta)
.
\end{align}
\end{subequations}
Examining the $\bq\rightarrow 0$ limit of $V^{ad}_{\al}(\bq)$ and $L^{-1}Z^{*}_{\al}(\bq)$ reveals that for both of them, the leading order term is $\mathcal{O}(\bq)$. This becomes clear from the $\bq\rightarrow 0$ limit, given in Appendix~\ref{mat_coef_app}, of the constituent variables $\mu_{\al}(\bq),\nu(\bq),\tau_{\al}(\bq),\theta(\bq)$ of these two terms. From Eq.~\eqref{heat_transport}c and Eq.~\eqref{defect_diff} we know that time derivatives of $\delta q$ and $\delta c$ has leading order $\mathcal{O}(\bq^{2})$ contributions. Therefore, time derivative of the momentum density equation Eq.~\eqref{EOM_j_adia} yields $\mathcal{O}(\bq^{2})$ contributions from the first term while the $\bq-$dependence is of the order $\mathcal{O}(\bq^{3})$ for the terms associated with $\partial_{t}\delta q(\bq,t)$ and $\partial_{t}\delta c(\bq,t)$. Ignoring the terms with higher order $\bq-$dependence, after substituting $\partial_{t}\delta u_{\al}(\bq,t)$ with Eq.~\eqref{EOM_olt}d, we arrive at the wave equation
\begin{align}\label{wave_eq}
    \partial^{2}_{t}\delta j_{\al}(\bq,t)=-n^{-1}_{0}\Lambda^{ad}_{\albe}(\bq)\delta j_{\be}(\bq,t).
\end{align}
The matrix $\Lambda^{ad}_{\albe}(\bq)$, defined in Eq.~\eqref{lam_V}a, is the adiabatic dynamical matrix~\cite{Wallace1998} for the crystalline solid. It is associated with all the adiabatic elastic coefficients and governs the longitudinal and transverse speeds of sound in the crystalline solid under adiabatic conditions where heat and local defects can adjust freely.

In defining the viscosity tensor $\eta_{\albe\gamma\delta}$ (Eq.\ref{ons_coef_set2}) we have not decomposed the longitudinal and transverse contributions of the stress fluctuations. In case of a simple fluid, the attenuation of the propagative sound waves associated with longitudinal components of the current correlation functions~\cite{BoonYip}, is shown~\cite{Forster} to be related to the viscosity tensor. Components of the viscosity tensor are also shown to govern the diffusion of the transverse shear waves associated with the transverse components of the current correlation functions. There, the reactive part of the time evolution equation of the momentum density is given by the pressure gradient alone, the second term in Eq.~\eqref{EOM_olt}c. Our calculations reveal that, in case of a crystalline solid, an additional reactive contribution arises in the equation of motion of the momentum density. From the literature~\cite{Fleming_Cohen,Mabi_1} on the macroscopic hydrodynamics of crystalline solids it is well known that this indicates the emergence of propagating shear waves. This marks another important difference between a system of simple fluid and a solid with long range order. A completely microscopic definition of the viscosity tensor proves to be another important result of this paper.

\section{Conclusions and Outlook}\label{conclusion}
Mechanical properties and transport coefficients, in a crystalline phase with a finite concentration of point-defects, are derived in this paper from a completely microscopic classical description. The hydrodynamic equations of the three components of the displacement vectors and the variables governing the local conservation of mass, energy, and linear momentum are derived. Our results are based on the Zwanzig-Mori formalism, which requires a choice of the considered variables, and on an ansatz for the microscopic density fluctuations in terms of the coarse-grained fields, Eq.~\eqref{ansatz1}. The choice of the relevant variables is dictated by conservation laws and spontaneous symmetry breaking. The ansatz for the displacement field (Eq.~\eqref{ansatz3_sum_b}), first suggested by Szamel and Ernst \cite{Szamel_Ernst_1993} has been tested in hard spheres \cite{lin2021} and cluster crystals~\cite{SG_jcp_2022}.
How this hydrodynamic description relates to the equilibrium thermodynamic properties like free energy and entropy was rigorously  worked out too. Our approach allows us to derive the transport coefficients associated with the Green-Kubo relations. The rank of the transport coefficient tensors reflect the crystal symmetries and their explicit microscopic origin provides the possibility of evaluating them from inputs obtained from atomistic simulations of appropriate systems. 
Theoretical and simulation studies in reference~\cite{SG_jcp_2022} show how theoretical frameworks derived in references~\cite{Walz2010,Haring2015} can be successfully implemented to obtain quantitative insights regarding mechanical response of solids. Though the previous studies~\cite{Haring2015,SG_jcp_2022} were restricted to reversible mechanical response in isothermal crystals, the theories success in predicting elastic properties of a model known to mimic crystalline phases in DNA-based dendritic nanostructures~\cite{DNA_CC}, illustrates the scope of applicability of this formulation. 
%\com{Another recent contribution~\cite{haussmann} by one of the authors, extends  this through the use of projection operators defined~\cite{Kawasaki1973} for ensembles far from equilibrium but with an assumption of local entropy maximisation.}

This paper provides a complete microscopic formulation of all the constants of linear elasticity and coefficients governing diffusive and wave transport processes for crystals. It has been established~\cite{Haring2015,SG_jcp_2022,lin2021}, albeit in a much more simplified purely reversible version of this theory, that the predictions can be validated using inputs from atomistic simulations of crystalline phases of particles interacting via ultra-soft potentials~\cite{MladekJPCB2007,DNA_CC}. Recent advances in experimental techniques allow the tracking of single particle dynamics in a thermodynamic ensemble of soft materials~\cite{colloid_rev1,colloid_rev2}. This offers the exiting possibility of understanding the emergent macroscopic properties from experimentally observable microscopic dynamics through the lens of a statistical mechanics theory derived from first principles. 

From a macroscopic phenomenology perspective, heat transport is an irreversible process related to the production of entropy in the system. Reference ~\cite{pre1_GB}, a study of meta-stable polycrystals by some of the authors, presents a completely different point of view centered around the relation of entropy to statistical microstates. In this case our extensive atomistic simulations give us insights about the relative stability of topological-defect rich polycrystalline configurations and allows us to discern the role of entropy in it. Therefore, combining the theoretical phenomenology of this paper and its predecessors ~\cite{Walz2010,Haring2015,SG_jcp_2022} with simulations similar to ~\cite{pre1_GB}, can lead to fundamentally new ways of interpreting reversible or irreversible thermo-mechanical response of crystalline solids and understanding transport processes driven by gradients in intensive thermodynamic fields like temperature.

%Phenomenological or atomistic theories attempting to explain macroscopic properties of crystals with defects argue~\cite{Fleming_Cohen,Mladek2007} for additional thermodynamic variables to describe an ensemble of particles with long range periodicity. 
Defining the displacement fields unambiguously, in a system with long range periodicity and diffusing defects, proves to be one of the conceptual challengers to phenomenological or atomistic theories attempting to explain macroscopic mechanical response~\cite{sethna}. Our approach of representing the system in reciprocal space, resolves this problem by defining displacements, particle number density or defect density in terms of density fluctuations close to Bragg-peaks. A natural extension of this would be to examine the possibility of employing similar principles to study systems subjected to large deformations near the onset of plasticity.

Some recent research employs spatial projection operators~\cite{sas1,sas2,popli} to segregate microscopic displacement fluctuations associated with macroscopic elastic or plastic response in defect-free crystals. This interpretation, of displacement fields, helps in explaining~\cite{saspnas} the origin of rigidity and the shear rate dependence of the yield-point~\cite{theoprl} in an ideal crystal. In this paper, the projection operators derive the dynamics of variables chosen because of their slow relaxation time-scales. The propagating longitudinal or transverse sound modes, described here, exhibit dispersion relations that vanish linearly with decreasing wave-vector. The slopes of their dispersion relations in the small wave-number limit also give the moduli of linear elastic response encoded in the dynamical matrix appearing in the wave equation of the displacement fields~\cite{Walz2010}. The spatial projection of reference~\cite{sas1} may present a way to separate the contribution of affine macroscopic deformations and local particle motions to the total microscopic displacement fluctuations in a defect rich crystal. Therefore, one of the future avenues for investigation will be an attempt to understand the onset of plasticity through the convergence of these perspectives. 

Identifying connections, between the current theoretical framework and reversible or irreversible isothermal and adiabatic processes in the crystal with local defects, will pave the path for future endeavours to evaluate macroscopic mechanical constants and transport coefficients of materials of theoretical and practical interest.

%\begin{itemize}
 %   \item Reiterate the importance of a theory framework capable of handling local-defects while describing linear elasticity
  %  \item How the simulations performed are exactly how actual experiments are performed while measuring elastic moduli. Contrast this to our theory approach as well as Mladek simulations
   % \item Reiterate the distinctions to hard-sphere systems
    %\item The importance of the localisation parameter and its influence on the mechanical response
    %\item Understanding linear elasticity, eventual understanding of non-linear and plastic response 
%\end{itemize}

\begin{acknowledgments}
Discussions with Gerhard Kahl, Martin Oettel, Thomas Franosch, and Johannes H\"aring are gratefully acknowledged.
This work is supported by Deutsche Forschungsgemeinschaft through  grant FU 309/11-1.
%\com{mf: please check whether we cite all papers by Gaurav, Kahl and Oettel that came out of the DACH grant, and check whether they mentioned our grant FU... No, I found that they didn't do it as it would be confusing to mention a grant to somebody not coauthor; thus i suggest to drop the next lines.}
%and OE 285/5-1, and the Austrian Funding Agency (FWF) under grant number I3846-N36. 
\end{acknowledgments}

%\section*{Data Availability Statement}

%The data that support the findings of this study are available within the article.

\appendix

\section{Some conventions and definitions}
\subsection{The microscopic variables and their Fourier transforms}\label{def_micro_app}
Assuming an interaction potential $V(|{\bf r}_{i}-{\bf r}_{j}|)=V(r_{ij})$ dependent on the distance $\br_{ij}$ between particles $i$ and $j$, in Fourier space, the microscopic stress tensor and the microscopic energy current reads,
\begin{widetext}
\begin{equation}\label{stress_app}
\hat{\boldsymbol{\sigma}}(\mathbf{q},t) = - \sum_i \frac{\hat{\bf p}(\br_{i},t) \hat{\bf p}(\br_{i},t)}{m} e^{-i \mathbf{q} \cdot \mathbf{r}_{i}(t)} + \frac{1}{2} \sum_{i\neq j} \frac{\mathbf{r}_{ij}(t)\mathbf{r}_{ij}(t) }{r_{ij}(t)} V'(r_{ij}(t))\frac{e^{-i \mathbf{q} \cdot \mathbf{r}_{j}(t)} - e^{-i \mathbf{q} \cdot \mathbf{r}_{i}(t)}}{i \mathbf{q} \cdot \mathbf{r}_{ij}(t)},
\end{equation}

\begin{align}\label{energy_curr_app}
\hat{\mathbf{j}}^e(\mathbf{q},t) &= \sum_i E(\br_{i},t) \frac{\hat{\mathbf{p}}(\br_{i},t)}{m} e^{-i \mathbf{q} \cdot \mathbf{r}_{i}(t)} - \frac{1}{4} \sum_{i \neq j} \frac{\hat{\mathbf{p}}(\br_{i},t) + \hat{\mathbf{p}}(\br_{j},t)}{m} \cdot \frac{\mathbf{r}_{ij}(t) \mathbf{r}_{ij}(t)}{r_{ij}(t)} V'(r_{ij}(t)) \frac{e^{-i \mathbf{q} \cdot \mathbf{r}_{j}(t)} - e^{-i \mathbf{q} \cdot \mathbf{r}_{i}(t)}}{i \mathbf{q} \cdot \mathbf{r}_{ij}(t)}
\end{align}
\end{widetext}

And one can indeed verify that \cite{stress_unique_95}
\begin{equation} 
\int d \mathbf{r} e^{-i \mathbf{q} \cdot \mathbf{r}} \int_0^1 d s \,\delta(\mathbf{r} - \mathbf{r}_{i} + s\mathbf{r}_{ij}) = \frac{e^{-i \mathbf{q} \cdot \mathbf{r}_{j}} - e^{-i \mathbf{q} \cdot \mathbf{r}_{i}}}{i \mathbf{q} \cdot \mathbf{r}_{ij}}.
\end{equation}
In the microscopic definitions for the energy density (Eq.~\eqref{en_micro}), stress (Eq.~\eqref{stress_app}) and energy current (Eq.~\eqref{energy_curr_app}), the per particle mass $m$ appears. This however, is set to one without any loss of generality and therefore it does not appear in any other equations in the rest of the paper.

\subsection{Equilibrium values}\label{app_e_p_eq}
We can define an equilibrium pressure, %\com{do we need it at finite q ? or just eq (A5)?  and is it lattice periodic?}
%\begin{align}
%p(\mathbf{q}) &= n(\mathbf{q}) \kT - \frac{1}{6} \langle \sum_{i\neq j} r_{ij} V'(r_{ij}) \frac{e^{-i \mathbf{q} \cdot \mathbf{r}_j}- e^{-i \mathbf{q} \cdot \mathbf{r}_i}}{i \mathbf{q} \cdot \mathbf{r}_{ij}} \rangle.
%\end{align}
using the virial equation for the average pressure
\begin{equation}
p_0 = n_0 \kT - \frac{1}{6V} \langle \sum_{i\neq j} r_{ij} V'(r_{ij}) \rangle.
\end{equation}

The last quantity that is not equally zero at equilibrium is the energy density,
\begin{align}
\langle \hat{e}(\mathbf{r}) \rangle &= \langle \sum_i E_i \delta(\mathbf{r} - \mathbf{r}_i) \rangle \nonumber\\
&= \langle \sum_i \frac{\hat{\mathbf{p}}_i^2}{2m} \delta(\mathbf{r} - \mathbf{r}_i) \rangle + \frac{1}{2} \langle \sum_{i\neq j} V(r_{ij}) \delta(\mathbf{r} - \mathbf{r}_i)\rangle \nonumber\\
&= \frac{3}{2} k_BT n(\mathbf{r}) + \frac{1}{2} \langle \sum_{i\neq j} V(r_{ij}) \delta(\mathbf{r} - \mathbf{r}_i)\rangle.
\end{align}

Non-locally, taking an ensemble and volume average, we get
\begin{equation}
e_0 = \frac{1}{V} \int d \mathbf{r} \, \scal{e(\mathbf{r})} \\ 
=\frac{3}{2}n_0 k_B T + \frac{1}{2V}\scal{\sum_{i\neq j} V(r_{ij})}.
\end{equation}
In reciprocal space, they translate to %\com{again, do we need the average at finite q and would it be lattice periodic?}
%	\begin{equation}\label{en_micro_app}
%	\scal{\hat{e}(\mathbf{q})} = \scal{\sum_i E_i e^{- i \mathbf{q} \cdot \mathbf{r}i)}} \\
%	= \frac{3}{2} k_BT n(\mathbf{q}) + \frac{1}{2} \scal{ \sum_{i\neq j} V(r_{ij})  e^{- i \mathbf{q}  \cdot \mathbf{r}_{i}}}.
%	\end{equation}
	\begin{equation}
	e_0 = \frac{1}{V} \scal{\hat{e}(\mathbf{q}=0)} \\ 
	=\frac{3}{2}n_0 k_B T + \frac{1}{2V}\scal{\sum_{i\neq j} V(r_{ij})}.
	\end{equation}

\subsection{Matrix components}\label{app_mat_cor}
The momentum-momentum density correlation is straightforward with the equipartition principle,
\begin{equation}
\scal{\delta \hat{\mathbf{j}}^*(\mathbf{q}) \delta \hat{\mathbf{j}}(\mathbf{q})} = n_{0}k_{B}TV \mathbb{I}
\end{equation}
With $\liou = -i \frac{d}{d t}$ and Eq.\ref{ceq_ng}, we get the first element of the frequency matrix
\begin{subequations}\label{omega_jrho}
	\begin{align}
	&\bm{\omega}^{j\rho}_{\bg\al}=\beta\scal{\delta \hat{j}^{*}_{\al}(\mathbf{q}) \liou \delta \hat{\rho}_{\bg}(\mathbf{q})}=-V(g+q)_{\al}n_{\bg}\\
	&\bm{\omega}^{\rho j}_{\bg\al}=\beta\scal{\delta \hat{j}^{*}_{\al}(\mathbf{q}) \liou \delta \hat{\rho}_{\bg}(\mathbf{q})}^{*}=-V(g+q)_{\al}n^{*}_{\bg}
	\end{align}
\end{subequations}
The second one reads
\begin{widetext}
\begin{subequations}\label{omega_je}
	\begin{align}
	&\beta^{-1}\bm{\omega}^{je}=\beta^{-1}\bm{\omega}^{ej}=\scal{\delta \bj^*(\mathbf{q}) \liou \de(\mathbf{q})} = -\bq \cdot \scal{\delta \hat{\mathbf{j}}^*(\mathbf{q}) \hat{\mathbf{j}}^e(\mathbf{q})} \\
	&= -\mathbf{q} \cdot \scal{ \sum_{i,k} E(\br_{i}) \frac{\hat{\mathbf{p}}_{i}}{m} \hat{\mathbf{p}}_{k}  e^{-i \mathbf{q} \cdot \mathbf{r}_{ik}} }+ \mathbf{q} \cdot \scal{\frac{1}{4} \sum_k \sum_{i \neq j}  \frac{\hat{\mathbf{p}}_{i} + \hat{\mathbf{p}}_{j}}{m} \cdot \frac{\mathbf{r}_{ij} \mathbf{r}_{ij}}{r_{ij}} \hat{\mathbf{p}}_{k} V'(r_{ij}) \frac{e^{-i \mathbf{q} \cdot \mathbf{r}_{j}} - e^{-i \mathbf{q} \cdot \mathbf{r}_{i}}}{i \mathbf{q} \cdot \mathbf{r}_{ij}}e^{i \mathbf{q} \cdot \mathbf{r}_{k}}} \\
	&= -\mathbf{q} \cdot \scal{ \sum_{i,k}  \frac{\hat{\mathbf{p}}_{i}^2}{2m}  \frac{\hat{\mathbf{p}}_{i}}{m} \hat{\mathbf{p}}_{k} e^{-i \mathbf{q} \cdot \mathbf{r}_{ik}} } -  \mathbf{q} \cdot \scal{ \sum_k \sum_{i\neq j} V(r_{ij}) \frac{\hat{\mathbf{p}}_{i}}{2m} \hat{\mathbf{p}}_{k}e^{-i \mathbf{q} \cdot \mathbf{r}_{ik}} }  + \mathbf{q} \cdot \scal{\frac{k_BT}{2} \sum_{i \neq j} \frac{\mathbf{r}_{ij}\mathbf{r}_{ij}}{r_{ij}} V'(r_{ij}) \frac{\sin(\mathbf{q} \cdot \mathbf{r}_{ij})}{\mathbf{q} \cdot \mathbf{r}_{ij}}} \\
	&= -\mathbf{q} \frac{5}{2} N \left(k_BT \right)^2 -  \mathbf{q} \frac{1}{2} k_BT \scal{\sum_{i\neq j} V(r_{ij}) }   + \mathbf{q} \frac{k_BT}{6} \scal{\sum_{i \neq j}r_{ij} V'(r_{ij})} + \mathcal O(\mathbf{q}^3) \\
	&= -\mathbf{q} k_BTV \left( e_0 + p_0 \right) + \mathcal O(\mathbf{q}^3)
	\end{align}
	\end{subequations}
\end{widetext}
%	\com{this is the place where averages over energy and pressure only show up, or? then $q=0$ is enough}
with at equilibrium $\ra \rb = \dfrac{1}{3} r^2 \delta_{\alpha\beta}$, and
\begin{align}
&\scal{\hat{\mathbf{p}}_{i} \hat{\mathbf{p}}_{k}} = m k_{B}T  \delta_{ik} \mathbb{I} ,\ \ 
\scal{\hat{\mathbf{p}}_{i}^2} = 3m k_{B}T ,\nonumber\\
&\scal{\hat{\mathbf{p}}_{i}^2  \hat{\mathbf{p}}_{i} \hat{\mathbf{p}}_{k}} = 5 \left( m k_BT \right)^2 \delta_{ik} \mathbb{I} ,\nonumber\\
&\scal{\hat{\mathbf{p}}_{i}^2  \hat{\mathbf{p}}_{k} \hat{\mathbf{p}}_{k}} = \left(3 + 2 \delta_{ik} \right)  \left( m k_BT \right)^2 \mathbb{I}.
\end{align}

\section{Inversion of the static susceptibility matrix}\label{chi_inv}
In the following matrix identities, the matrix blocks $\bm{A,B,C,D}$ are square matrices. We use these identities (Eq.~\eqref{Mat_id}) to perform the inversion of the static susceptibility matrix $\bm{\chi}$ (Eq.~\eqref{stat_cor_appendix}).
\begin{widetext}
%\begin{subequations}
\begin{align}\label{Mat_id}
    &
    \begin{bmatrix}
    \bm{A} & 0\\
    0&\bm{B}
    \end{bmatrix}^{-1}
    =
    \begin{bmatrix}
    \bm{A}^{-1} & 0\\
    0 & \bm{B}^{-1}
    \end{bmatrix}, \ \ 
    \begin{bmatrix}
    \bm{A}& \bm{B}\\
    \bm{C}&\bm{D}
    \end{bmatrix}^{-1}
    =
    \begin{bmatrix}
    \bm{A}^{-1}+\bm{A}^{-1}\bm{B}\left(\bm{D}-\bm{C}\bm{A}^{-1}\bm{B}\right)^{-1}\bm{C}\bm{A}^{-1} & -\bm{A}^{-1}\bm{B}\left(\bm{D}-\bm{C}\bm{A}^{-1}\bm{B}\right)^{-1}\\
    -\left(\bm{D}-\bm{C}\bm{A}^{-1}\bm{B}\right)^{-1}\bm{C}\bm{A}^{-1} & \left(\bm{D}-\bm{C}\bm{A}^{-1}\bm{B}\right)^{-1}
    \end{bmatrix}
\end{align}
%\end{subequations}
\begin{align}\label{stat_cor_appendix}
\bm{\chi}(\bq)&=
\begin{bmatrix}
\bm{\chi}^{\rho\rho}_{(N\times N)} & \bm{\chi}^{\rho e}_{(N\times 1)} &\bm{\chi}^{\rho j}_{(N\times 3)} \\
\bm{\chi}^{e\rho}_{(1\times N)} & \bm{\chi}^{e e}_{(1\times 1)} &\bm{\chi}^{e j}_{(1\times 3)} \\
\bm{\chi}^{j\rho}_{(3\times N)} & \bm{\chi}^{je}_{(3\times 1)} & \bm{\chi}^{j j}_{(3\times 3)}
\end{bmatrix}=\begin{bmatrix}
\bm{\chi}^{\rho\rho}_{(N\times N)} & \bm{\chi}^{\rho e}_{(N\times 1)} & 0 \\
\bm{\chi}^{e\rho}_{(1\times N)} & \bm{\chi}^{e e}_{(1\times 1)} & 0 \\
0 & 0 & \bm{\chi}^{j j}_{(3\times 3)}
\end{bmatrix}
\end{align}
\end{widetext}

Reiterating Eq.~\eqref{stat_cor0} and noting the block diagonal structure of $\bm{\chi}$ (Eq.~\eqref{stat_cor_appendix})
first the individual diagonal blocks are inverted using identity Eq.~\eqref{Mat_id}a. The first diagonal block, comprising of correlation matrices $\bm{\chi}^{\rho\rho}$, $\bm{\chi}^{\rho e}$ and $\bm{\chi}^{ee}$ is inverted using identity Eq.~\eqref{Mat_id}b. The second diagonal block, with correlation between components of  momentum densities, is easier to invert using the classical equipartition (Eq.~\eqref{cl_eq_part}). Thus the explicit expressions for the inverse matrix
\begin{align}\label{stat_cor1}
\chi^{-1}(\bq) = 
\begin{bmatrix}
({\bf J}^{\rho\rho} + {\bf U L^{-1} U^{*}}) & - {\bf L^{-1}U} & 0 \\
- ({\bf L^{-1} U)^{*}} & {\bf L^{-1}} & 0 \\
0 & 0 & n_0^{-1} \mathbb{I}.
\end{bmatrix}
\end{align}
is acquired where ${\bf J}^{\rho\rho}=(\bm{\chi}^{\rho\rho})^{-1}$ with components $J_{\bg\bg'}$ (Eq.~\eqref{jgg}). The scalar $L(q)$ and the components $U_{\bg}$ of the vector $U$ are given in Eq.~\eqref{KLU_micro}.
%and repeated here.
%\begin{subequations}\label{KLU_micro_app}
%\begin{align}
%    &K_{\bg}(\bq)=\beta\langle \delta e^{*}(\bq)\delta n_{\bg}(\bq)\rangle\\
%    &K(\bq)=\beta\langle \delta e^{*}(\bq)\delta e(\bq)\rangle\\
%    &L(\bq)=K(\bq)-\sum_{\bg\bg'}K_{\bg}(\bq)J_{\bg\bg'}(\bq)K^{*}_{\bg'}(\bq)\\
%    &U_{\bg}(\bq)=\sum_{\bg'}J^{*}_{\bg\bg'}(\bq)K_{\bg'}(\bq)
%\end{align}
%\end{subequations}

\section{Small wavelength limit of coefficients related to the inverse density correlation function and hence the direct correlation function}\label{mat_coef_app}	
	
\subsection{The generalised elastic coefficients}\label{elastic_cons_app}
The elastic coefficients, which are the same as the ones in ~\cite{Walz2010,SG_jcp_2022}, are summarised  %\com{I would just give the integrals in C1 and the final integrals for the greeks including the direct cor. fnct.}
\begin{subequations}\label{lam_mu_nu_app}
\begin{align}
&\lambda_{\alpha\beta}(\bq)=\sum_{\bg'\bg''}g'_{\alpha}n^{*}_{\bg'}J^{*\rho\rho}_{\bg'\bg''}n_{\bg''}g''_{\beta}=\lambda_{\alpha\beta\gamma\delta}q_{\gamma}q_{\delta}+\dots\\
&\mu_{\alpha}(\bq)=\sum_{\bg'\bg''}n^{*}_{\bg'}J^{*\rho\rho}_{\bg'\bg''}n_{\bg''}ig''_{\alpha}=i\mu_{\alpha\beta}q_{\beta}+\dots\\
&\mu^{*}_{\alpha}(\bq)=\sum_{\bg'\bg''}-ig'_{\alpha}n^{*}_{\bg'}J^{*\rho\rho}_{\bg'\bg''}n_{\bg''}=-i\mu_{\alpha\beta}q_{\beta}+\dots\\
&\nu(\bq)=\sum_{\bg'\bg''}n^{*}_{\bg'}J^{*\rho\rho}_{\bg'\bg''}n_{\bg''}=\nu+\dots
\end{align}
\end{subequations}
and then derived here for the sake of completeness.

 Substituting $J^{*}_{\bg'\bg}$ using Eq.~\eqref{jgg} in the expression for $\lambda_{\al\be}(\bf q)$ in Eq.~\eqref{lam_mu_nu_app}a and utilising the expansion of the gradient of the average density distribution 
\begin{equation}
    \nabla_{\alpha}n(\br)=\sum_{\bg}ig_{\alpha}n_{\bg}e^{i\bg.\br}\label{gradft}
\end{equation}
in terms of the Bragg peak amplitudes $n_{\bg}$, one obtains
%ends up getting the following expression for $\lambda_{\al\be}$
%
\begin{align}
\lambda_{\al \be}(\bq)
%&=-\sum_{\bg,\bg^{'}}ig^{'}_{\al}n^{*}_{\bg^{'}}J^{*}_{\bg^{'}\bg}(\bq)n_{\bg}ig_{\be}\\
%&=-\frac{k_{B}T}{V}\sum_{\bg,\bg^{'}} \int d^{3}r_{1}\int d^{3}r_{2}ig^{'}_{\al}n^{*}_{\bg^{'}} e^{-i\bg^{'}.\br_{2}}e^{-i\bq.(\br_{1}-\br_{2})}\left[\frac{\delta(\br_{1}-\br_2)}{n(\br_{1})}-c(\br_{1},\br_{2})\right]n_{\bg}ig_{\be}e^{i\bg.\br_{1}}, ~\text{using (\ref{jgg})}\\
=&\frac{k_{B}T}{V}\int d^{3}r_{1}\int d^{3}r_{2}\nabla_{\al}n(\br_{1})\nabla_{\beta}n(\br_{2})e^{-i\bq.(\br_{1}-\br_{2})}\nonumber\\
&\left[\frac{\delta(\br_{1}-\br_2)}{n(\br_{1})}-c(\br_{1},\br_{2})\right] .
\end{align}
Upon using an equation derived by Lovett, Mou, Buff, Wertheim (LMB~\cite{LMBW1}W~\cite{LMBW2}), 
\begin{equation}
    \frac{\nabla_{\al}(n(\br))}{n(\br)}=\int d^{3}r'c(\br,\br')\nabla_{\al}n(\br')\label{LMBW}
\end{equation}
and realising that the gradient of the equilibrium density $\nabla_{\al}n(\br)$ is real i.e., $\sum_{\bg}ig_{\alpha}n_{\bg}e^{i\bg.\br}=\sum_{\bg}-ig_{\alpha}n^{*}_{\bg}e^{-i\bg.\br}$, one gets
\begin{subequations}
\begin{align}
    \lambda_{\al\be}(\bq)=&\frac{k_{B}T}{V}\int d^{3}r_{1}\int d^{3}r_{2}\nabla_{\al}n(\br_{1})\nabla_{\beta}n(\br_{2})c(\br_{1},\br_{2})\nonumber\\
    &\left(1-e^{-i\bq\cdot(\br_{1}-\br_{2})}\right)\\
    \approx&\lambda_{\al\be\gamma\delta}q_{\gamma}q_{\delta}+\mathcal{O}(q^{4})
%&=-\frac{k_{B}T}{V}\sum_{\bg,\bg^{'}} \int d^{3}r_{1}\int d^{3}r_{2}ig^{'}_{\al}n^{*}_{\bg^{'}} e^{-i\bg^{'}.\br_{2}}n_{\bg}ig_{\be}e^{i\bg.\br_{1}}c(\br_{1},\br_{2})\left(1-e^{-i\bq.(\br_{1}-\br_{2})}\right)\\
\end{align}
\end{subequations}
Similar arguments lead to the expression for 
\begin{subequations}
\begin{align}
    \mu_{\al}(\bq)=&\frac{k_{B}T}{V}\int d^{3}r_{1}\int d^{3}r_{2}n(\br_{1})\nabla_{\al}n(\br_{2})c(\br_{1},\br_{2})\nonumber\\
    &\left(1-e^{-i\bq\cdot(\br_{1}-\br_{2})}\right)\\
    \approx&i\mu_{\al\be}q_{\be}+\mathcal{O}(q^{2})
\end{align}
\end{subequations}
It can, however, be shown~\cite{SG_jcp_2022} that for a crystal with inversion symmetry, the correction in the small $\bq$ expansion of $\mu_{\al}(\bq)$ is $\mathcal{O}(q^{3})$. Finally, the generalised elastic coefficient $\nu(\bq)$, whose leading order contribution comes from the homogeneous constant $\nu$, is given by
\begin{subequations}\label{nu_small_q_app}
\begin{align}
    \nu(\bq)=&\frac{k_{B}T}{V}\int d^{3}r_{1}\int d^{3}r_{2}n(\br_{1})n(\br_{2})e^{-i\bq.(\br_{1}-\br_{2})}\nonumber\\
&\left[\frac{\delta(\br_{1}-\br_2)}{n(\br_{1})}-c(\br_{1},\br_{2})\right]\\
\approx&\nu+\mathcal{O}(q^{2})
\end{align}
\end{subequations}
It can be shown~\cite{Walz2010} that, as a consequence of the $\br_{1}\leftrightarrow \br_{2}$ symmetry, $\nu(\bq)$ is real and has contributions from even powers in a long wavelength expansion in $\bq$.

\subsection{The coefficients
coupling to energy}\label{co_tau_theta_L_app}
In this section we derive the small wave vector limit for the coefficients $\tau_{\al},\theta$ and $L$. The initial definition of $\tau_{\al}$ (Eq.~\eqref{theta_tau_micro})

\begin{align}
    \tau_{\al}(\bq)=i\sum_{\bg}U^{*}_{\bg}(\bq)n_{\bq}g_{\al}
\end{align}
uses the abbreviated notation of $U_{\bg}$ defined in terms of the inverse density correlation function $J_{\bg\bg'}$ (see Eq.~\eqref{jgg}) and the correlation between fluctuation in energy and Bragg peak amplitudes $K_{\bg}$ (see Eq.~\eqref{KLU_micro}). Therefore, plugging in these definitions, $\tau_{\al}$ can be written in terms of the direct correlation function $c(\br_{1},\br_{2})$ and symmetry of these function can be exploited to derive an expression for $\tau_{\al}$ in the small $\bq$ limit.

\begin{widetext}
\begin{subequations}\label{tauab_app}
\begin{align}
     \tau_{\al}(\bq)&= (\beta V)^{-1}\sum_{\bg,\bg'}\int \dd^{3}r_{1}\int \dd^{3}r_{2}in_{\bg}g_{\al}e^{i\bg\cdot \br_{1}}K^{*}_{\bg'}(\bq)e^{-i\bg'\cdot\br_{2}}e^{-i\bq\cdot(\br_{1}-\br_{2})}\left[\dfrac{\delta (\br_{1}-\br_{2})}{n(\br_{1})}-c(\br_{1},\br_{2})\right]\\
     &=(\beta V)^{-1}\sum_{\bg'}\int \dd^{3}r_{2}\nabla_{\al}n(\br_{1})K^{*}_{\bg'}(\bq)e^{-i\bg'\cdot\br_{2}}e^{-i\bq\cdot(\br_{1}-\br_{2})}\left[\dfrac{\delta (\br_{1}-\br_{2})}{n(\br_{1})}-c(\br_{1},\br_{2})\right]\\
     &=(\beta V)^{-1}\sum_{\bg'}\int \dd^{3}r_{2}K^{*}_{\bg'}(\bq)e^{-i\bg'\br_{2}}\int \dd^{3}r_{1}\nabla_{\al}n(\br_{1})c(\br_{1},\br_{2})\left(1-e^{-i\bq\cdot (\br_{1}-\br_{2})}\right)\\
     &=V^{-2}\sum_{\bg'}\int \dd^{3}r_{2}\langle\sum_{jk}E_{j}e^{-i\bq\cdot (\br_{j}-\br_{k})}e^{i\bg'\cdot \br_{k}} \rangle e^{-i\bg'\br_{2}}\int \dd^{3}r_{1}\nabla_{\al}n(\br_{1})c(\br_{1},\br_{2})[1-i\bq.\br_{12}+\dots]\\
     &=iq_{\be}\tau_{\albe}+\mathcal{O}(q^{2})
\end{align}
\end{subequations}
where the second rank tensor $\tau_{\al\be}$ is
\begin{align}
\tau_{\albe}&=(\beta V)^{-1}\sum_{\bg'}\int \dd^{3}r_{2}\bar{K}^{*}_{\bg'} e^{-i\bg'\br_{2}}\int \dd^{3}r_{1}\nabla_{\al}n(\br_{1})c(\br_{1},\br_{2})r_{12,\beta}   
\end{align}
\end{widetext}
%\com{shouldn't we give the final integral for the new greek, $\tau_{\alpha\beta}$ ?}
Note the analogy between the definitions of $\mu_{\al}$ and $\tau_{\al}$. From this analogy, symmetry arguments applicable for $\mu_{\al\be}$ (Eq.~\eqref{lam_mu_nu_app}) (also see reference~\cite{Walz2010}), holds for $\tau_{\albe}$ as well. The symmetry $c(\br_{1},\br_{2})=c(\br_{2},\br_{1})$ and the LMBW equation indicates $\tau_{\albe}=\tau_{\be\al}$. The term $K^{*}_{\bg'}(\bq)$, in Eq.~\eqref{tauab_app}, is a $\bq$ dependent correlation function (see Eq.~\eqref{KLU_micro}a) between the energy density and Bragg diffraction amplitudes. In the $\bq\rightarrow 0$ limit the leading order term, in the expansion of the $\bq$ dependent exponential in its expression, is a constant $\bar{K}^{*}_{\bg'}$ and $\bq$ independent macroscopic property of the system. Similar to the correlations in Eq.~\eqref{omega_jrho}, this quantity is expected to have the periodicity of the lattice structure.

Next we take up the coefficient $\theta$. While deriving the explicit expressions for $\theta$ in terms of $c(\br_{1},\br_{2})$ we draw attention to the analogy between the definitions of $\theta$ and $\nu$.
\begin{widetext}
\begin{subequations}
\begin{align}
    \theta(\bq)&=\sum_{\bg}U^{*}_{\bg}(\bq) n_{\bg}=\sum_{\bg,\bg'}K^{*}_{\bg'}(\bq)J^{*}_{\bg'\bg}(\bq)n_{\bg}\\
    &=(\beta V)^{-1}\sum_{\bg,\bg'}\int \int \dd^{3}r_{1}\dd^{3}r_{2} n_{\bg}e^{i\bg\cdot \br_{1}}\bar{K}^{*}_{\bg'}e^{-i\bg'\cdot\br_{2}}e^{-i\bq\cdot(\br_{1}-\br_{2})}\left[\dfrac{\delta (\br_{1}-\br_{2})}{n(\br_{1})}-c(\br_{1},\br_{2})\right]\\
    &=(\beta V)^{-1}\sum_{\bg'}\int \int \dd^{3}r_{1}\dd^{3}r_{2} n(\br_{1})\bar{K}^{*}_{\bg'}e^{-i\bg'\cdot\br_{2}}e^{-i\bq\cdot(\br_{1}-\br_{2})}\left[\dfrac{\delta (\br_{1}-\br_{2})}{n(\br_{1})}-c(\br_{1},\br_{2})\right], \ \ \because n(\br_{1})=\sum_{\bg}n_{\bg}e^{i\bg\cdot \br_{1}}\\
    &=(\beta V)^{-1}\sum_{\bg'} \int\dd^{3}r_{2}\bar{K}^{*}_{\bg'}e^{-i\bg'\cdot\br_{2}}-(\beta V)^{-1}\sum_{\bg'}\int \int \dd^{3}r_{1}\dd^{3}r_{2} n(\br_{1})\bar{K}^{*}_{\bg'}e^{-i\bg'\cdot\br_{2}}c(\br_{1},\br_{2})[1-i\bq.\br_{12}+\dots]
\end{align}
\end{subequations}
The above equation shows how in the long wavelength limit, 
\begin{align}
\theta&=(\beta V)^{-1}\sum_{\bg'} \int\dd^{3}r_{2}\bar{K}_{\bg'}e^{-i\bg'\cdot\br_{2}}-(\beta V)^{-1}\sum_{\bg'}\int \int \dd^{3}r_{1}\dd^{3}r_{2} n(\br_{1})\bar{K}^{*}_{\bg'}e^{-i\bg'\cdot\br_{2}}c(\br_{1},\br_{2})   
\end{align}
\end{widetext}
is a $\bq$ independent real constant similar to $\nu$. Here, similar to $\tau_{\albe}$, we have used $\bar{K}^{*}_{\bg'}$ as the long wavelength expectation value for $K^{*}_{\bg'}(\bq)$. %\com{please give it explicitly}

Utilising the microscopic expressions for the energy fluctuations (see Eq.~\eqref{en_micro} and Eq.~\eqref{FT_e_op}) in the definition of $L$ (Eq.~\eqref{KLU_micro}c) and taking Taylor expansion of the exponential functions of $\bq$ leads to 
\begin{widetext}
\begin{subequations}
\begin{align}
    L(\bq)&=\beta \langle\delta \hat{e}^{*}\delta \hat{e}\rangle-\sum_{\bg,\bg'}K_{\bg}(\bq)J_{\bg\bg'}(\bq)K^{*}_{\bg'}(\bq) \\
    &=\beta \langle\delta \hat{e}^{*}\delta \hat{e}\rangle-(\beta V)^{-1}\sum_{\bg,\bg'}\int \int \dd^{3}r_{1}\dd^{3}r_{2} \bar{K}_{\bg}e^{i\bg\cdot \br_{1}}\bar{K}^{*}_{\bg'}e^{-i\bg'\cdot\br_{2}}\left[\dfrac{\delta (\br_{1}-\br_{2})}{n(\br_{1})}-c(\br_{1},\br_{2})\right]e^{-i\bq\cdot(\br_{1}-\br_{2})}\\
    &=\beta \langle\delta \hat{e}^{*}\delta \hat{e}\rangle-(\beta V)^{-1}\sum_{\bg,\bg'}\int \int \dd^{3}r_{1}\dd^{3}r_{2} \bar{K}_{\bg}e^{i\bg\cdot \br_{1}}\bar{K}^{*}_{\bg'}e^{-i\bg'\cdot\br_{2}}\left[\dfrac{\delta (\br_{1}-\br_{2})}{n(\br_{1})}-c(\br_{1},\br_{2})\right][1-i\bq.\br_{12}+\dots]
\end{align}
\end{subequations}
with the small wavelength limit of $L(\bq)$, given by a constant
\begin{align}
    L&=\beta \langle\delta \hat{e}^{*}\delta \hat{e}\rangle-(\beta V)^{-1}\sum_{\bg,\bg'}\int \int \dd^{3}r_{1}\dd^{3}r_{2} \bar{K}_{\bg}e^{i\bg\cdot \br_{1}}\bar{K}^{*}_{\bg'}e^{-i\bg'\cdot\br_{2}}\left[\dfrac{\delta (\br_{1}-\br_{2})}{n(\br_{1})}-c(\br_{1},\br_{2})\right]
\end{align}
\end{widetext}
%\com{again I would give the final integral, except for the energy variance}
Here, once again, in the $\bq\rightarrow 0$ limit, the leading order contributions from the terms $K_{\bg}(\bq)$ are $\bq$ independent constants like $\bar{K}_{\bg}$. The arguments related to the term $J_{\bg\bg'}(\bq)$, in the small $\bq$ limit, are identical to the ones given for the calculation of $\nu(\bq)$ (see Eq.~\eqref{nu_small_q_app} and reference~\cite{Walz2010,SG_jcp_2022,Flo_thesis}). %\com{What can we say about $\theta$ and $L$?, they are real, or? is $L>0 $? discuss arguments}

\section{coarse-grained static susceptibility}\label{sec_chi_cg_app}
This appendix aims to derive relations between (i) the intensive thermodynamic conjugate fields defined within the Mori-Zwanzig projection formalism and appearing on the left hand side of the Eq.~\eqref{rel_conjugates_cg}, 
%(also Eq.~\eqref{app_mat_conj_1} here) 
and (ii) the thermodynamic fields introduced in the expansion of the free energy in Eq.~\eqref{F_nuT_exp} and now used to represent the partition function in Eq.~\eqref{app_Z} (also appearing on the left hand side of Eq.~\eqref{app_conj_chi_nue}).

The relation between the coarse-grained thermodynamic variables in Eq.~\eqref{rel_conjugates_cg} is given by the static susceptibility matrix. 
%
% gives
% \begin{align}\label{app_mat_conj_1}
% &V^{-1}\bm{\chi}\begin{bmatrix}
%     \delta a  \\
%     \delta \bm{y}\\
%       \delta b\\
%     \delta \mathbf{v} \\  
% \end{bmatrix}
%     =
% \begin{bmatrix}
%   \delta n\\
%   \delta \bm{u}\\
%       \delta e\\
%   \delta \mathbf{j}
% \end{bmatrix}  . 
% \end{align}
%
The block diagonal structure of the coarse-grained susceptibility matrix $\bm{\chi}$, similar to its microscopic higher dimensional analogue in Eq.~\eqref{stat_cor0}, decouples a $3\times 3$ diagonal block of correlations between the linear momentum densities, $\bm{\chi^{jj}}$, from the $5\times 5$ matrix block $\bm{\chi_{nue}}$, allowing one to separately consider 
\begin{align}\label{app_mat_nue_conj_1}
&V^{-1}\bm{\chi_{nue}}\begin{bmatrix}
    \delta a  \\
    \delta \bm{y}\\
       \delta b
\end{bmatrix}
    =
\begin{bmatrix}
   \delta n\\
   \delta \bm{u}\\
      \delta e
\end{bmatrix}  . 
\end{align}
decoupled from the correlations to linear momentum. The matrix $\bm{\chi_{nue}}$ represents the static correlations between fluctuations in number density, displacement fields and energy density (see Eq.~\eqref{app_chi_macro_1}). In this appendix, we focus on interpreting the coarse-grained thermodynamic conjugate variables after defining the static correlation matrix $\bm{\chi_{nue}}$

\begin{widetext}
\begin{align}\label{app_chi_macro_1}
	\lim_{\bq \rightarrow 0} {\bm\chi_{nue}}(\bq) &= \lim_{\bq \rightarrow 0} \beta\left(
	\begin{array}{ccc}
	\scal{\delta n^{*}(\bq) \delta n(\bq)}& \scal{\delta n^{*}(\bq) \delta u_{\gamma}(\bq)} & \scal{\delta n^{*}(\bq) \de(\bq)}  \\
	\scal{\delta u_{\al}^{*}(\bq) \delta n(\bq)}& \scal{\delta u_{\al}^{*}(\bq) \delta u_{\gamma}(\bq)} & \scal{\delta u_{\al}^{*}(\bq) \de(\bq)}\\
	\scal{\de^{*}(\bq) \delta n(\bq)} &\scal{\de^{*}(\bq) \delta u_{\gamma}(\bq)} & \scal{\de^{*}(\bq) \de(\bq)} 
	\end{array}
	\right) 
	\end{align}
\end{widetext}
in terms of thermodynamic derivatives (see Eq.~\eqref{app_chi_macro_2}). These relations follow from evaluating the static correlations between the thermodynamic density fields (see Eq.~\eqref{app_LRT_conj_cor}) in the generalized grand-canonical ensemble  ~\cite{chaikin_lubensky_1995,Forster,Haring2015}. But for that, first we will have to define the partition function (see Eq.~\eqref{app_Z}).
In case of three dimensional systems, ${\bf u}$ is a three dimensional vector with components $u_{\al}$ corresponding to the three Cartesian coordinates. For ease of representation, in Eq.~\eqref{app_chi_macro_1}, we choose to show the correlations and the thermodynamic derivatives corresponding to one of the components of ${\bf u}$. These expressions are representative of more general susceptibility matrices with dimensions appropriate for the systems concerned. Reference ~\cite{Haring2015} explicitly derives these relations for an isothermal crystalline solid without considering heat transport associated with the fluctuations in energy density $\delta e$. For the isothermal case the coarse-grained susceptibility matrix, representing the correlations between the density $\delta n$ and displacement $\delta u_{\al}$ fluctuations, is a matrix of dimensions $4\times 4$. In this paper, with the additional energy fluctuations $\delta e$, the size of the static correlation matrix $\bm{\chi_{nue}}$ increases to $5\times 5$ to account for the additional thermodynamic correlations. Apart from this increase in the number of correlations involved, the main thermodynamic arguments remain identical.

Now we examine how the correlations in Eq.~\eqref{app_chi_macro_1} can be represented as thermodynamic derivatives obtained starting from the partition function. Drawing analogy to the statistical mechanics of ordered ferromagnetic states in a system with the Heisenberg Hamiltonian~\cite{chaikin_lubensky_1995}, we introduce the elastic energy ${\bf h}\cdot {\bf u}$ in the Hamiltonian $\mH$. Here we follow the definitions of intensive stress fields $h_{\albe}$ introduced as the thermodynamic conjugates to the symmetric linear strain fields $u_{\albe}$ in the free energy expansion in Eq.~\eqref{F_nuT_exp}. The partition function $\mathcal{Z}$ corresponds to the equilibrium crystalline solid in the limit of $h_{\al\be}\rightarrow 0$.
\begin{equation}\label{app_Z}
    \mathcal{Z}=\int \dd \Gamma e^{-\beta \mH+\beta\mu N-\beta V{\bf h}\cdot{\bf u} }
\end{equation}
Within the linear response picture~\cite{LL_ET}, the vector ${\bf u}$ has six components representing the Voigt symmetric strain fields $u_{\albe}$ : the independent components being for $\albe=11,22,33,23 \text{ or } 32,13 \text{ or } 31, 12 \text{ or } 21$. From a thermodynamics perspective, the components of the conjugate stress fields $h_{\al\be}$ can be defined as in Eq.~\eqref{maxw_rel}c where the coefficients of the stiffness tensor $C^{n}_{\albe\gamma\delta}$ are the constants connecting the conjugate pairs. For a thermodynamic variable density $w$ measured at $\bq\rightarrow 0$, the following relations for various density correlations can be derived from Eq.~\eqref{app_Z} \cite{chaikin_lubensky_1995} 

\begin{subequations}\label{app_LRT_conj_cor}
\begin{align}
 &\thermodevin{w}{(\beta \mu)}{\beta, \beta {\bf h}}=\langle \delta w^{*} \delta n\rangle\\
  -&\thermodevin{w}{(\beta h_{\albe})}{\beta,\beta \mu, \beta h_{\gamma\delta}}=\langle \delta w^{*} \delta u_{\albe}\rangle \\
 -&\thermodevin{w}{(\beta)}{\beta \mu, \beta {\bf h}}=\langle \delta w^{*} \delta e\rangle. 
\end{align}
\end{subequations}

For our system of interest, the thermodynamic variable density $w$ denotes number density $n$, energy density $e$ and linear symmetric strain fields $u_{\albe}$, which can be equivalently represented in terms of the displacement fields as given in Eq.~\eqref{def_sym_uab}. Consider representing fluctuations in these three quantities in terms of the partial derivatives of the three intensive fields $\beta \mu, \beta h_{\albe}$ and $\beta$ now introduced in Eq.~\eqref{app_LRT_conj_cor} through the definition of the partition function in Eq.~\eqref{app_Z}. With $\sum_{i=1}^{n}\dfrac{\partial f}{\partial x_{i}}\delta x_{i}=\delta f$ being a general form for the total derivative of a function $f$ of variables $x_{1}, x_{2},\dots, x_{n}$, the quantities $\delta n, \delta u_{\gamma\delta}$ and $\delta e$ can be written as follows

\begin{widetext}
\begin{subequations}\label{app_tot_der_nue_1}
\begin{align}
 &\thermodevin{n}{(\beta \mu)}{\beta, \beta {\bf h}}\delta (\beta \mu) - \thermodevin{n}{(\beta h_{\albe})}{\beta,\beta \mu, \beta h_{\gamma\delta}} \delta (\beta h_{\albe})-  \thermodevin{n}{(\beta)}{\beta \mu, \beta {\bf h}}\delta \beta=\delta n\\
&\thermodevin{u_{\gamma\delta}}{(\beta \mu)}{\beta, \beta {\bf h}}\delta (\beta \mu) - \thermodevin{u_{\gamma\delta}}{(\beta h_{\albe})}{\beta,\beta \mu, \beta h_{\gamma\delta}} \delta (\beta h_{\albe})-  \thermodevin{u_{\gamma\delta}}{(\beta)}{\beta \mu, \beta {\bf h}}\delta \beta=\delta u_{\gamma\delta}\\
&\thermodevin{e}{(\beta \mu)}{\beta, \beta {\bf h}}\delta (\beta \mu) - \thermodevin{e}{(\beta h_{\albe})}{\beta,\beta \mu, \beta h_{\gamma\delta}} \delta (\beta h_{\albe})-  \thermodevin{e}{(\beta)}{\beta \mu, \beta {\bf h}}\delta \beta=\delta e.
\end{align}
\end{subequations}
\end{widetext}
Einstein convention for summation over repeated indices has been used here. Since the linear symmetric strain fields $\delta u_{\albe}$ can be written in terms of the displacement fields (Eq.~\eqref{def_sym_uab}), we choose to follow the analogy of Eq.~\eqref{def_yab} to define the vector $\delta \bm{h}$ in terms of the stress fields $\delta h_{\albe}$ such that the components of $\delta \bm{h}$ 
\begin{align}\label{app_def_hab}
 &\delta h_{\al}=-i\delta h_{\albe}q_{\be},   
\end{align}
 are conjugates to $\delta u_{\al}$. When $\delta u_{\albe}$ and $\delta h_{\albe}$ are substituted with $\delta u_{\al}$ and $\delta h_{\al}$ respectively in Eq.~\ref{app_tot_der_nue_1}, they transform to 
 \begin{widetext}
 \begin{subequations}\label{app_tot_der_nue_2}
\begin{align}
&\thermodevin{n}{(\beta \mu)}{\beta, \beta {\bf h}}\delta (\beta \mu) - \thermodevin{n}{(\beta h_{\al})}{\beta,\beta \mu, \beta h_{\gamma}} \delta(\beta h_{\al})-  \thermodevin{n}{(\beta)}{\beta \mu, \beta {\bf h}}\delta \beta=\delta n\\
&\thermodevin{u_{\gamma}}{(\beta \mu)}{\beta, \beta {\bf h}}\delta (\beta \mu) - \thermodevin{u_{\gamma}}{(\beta h_{\al})}{\beta,\beta \mu, \beta h_{\gamma}} \delta(\beta h_{\al})-  \thermodevin{u_{\gamma}}{(\beta)}{\beta \mu, \beta {\bf h}}\delta \beta=\delta u_{\gamma}\\
&\thermodevin{e}{(\beta \mu)}{\beta, \beta {\bf h}}\delta (\beta \mu) - \thermodevin{e}{(\beta h_{\al})}{\beta,\beta \mu, \beta h_{\gamma}} \delta(\beta h_{\al})-  \thermodevin{e}{(\beta)}{\beta \mu, \beta {\bf h}}\delta \beta=\delta e.
\end{align}
\end{subequations}
 \end{widetext}
The set of equations in Eq.~\eqref{app_tot_der_nue_2} can be contracted into the matrix form
\begin{align}\label{app_conj_chi_nue}
&V^{-1}\bm{\chi_{nue}}\begin{bmatrix}
    \delta (\beta \mu)  \\
    \delta (\beta\bm{h})\\
       \delta \beta
\end{bmatrix}
    =
\begin{bmatrix}
   \delta n\\
   \delta \bm{u}\\
      \delta e
\end{bmatrix}. 
\end{align}
if the matrix $\bm{\chi_{nue}}$ is given by Eq.~\eqref{app_chi_macro_2a}. Next, recall Eq.~\eqref{app_LRT_conj_cor}, which gives linear response relations between static correlation functions and respective thermodynamic derivatives. This can now be used to obtain Eq.~\eqref{app_chi_macro_2b} from Eq.~\eqref{app_chi_macro_2a}.
\begin{widetext}
\begin{subequations}\label{app_chi_macro_2}
\begin{align}
	\lim_{\bq \rightarrow 0} {\bm\chi_{nue}}(\bq)
    &	= \beta V \left(
	\begin{array}{ccc}
	\thermodevin{n}{(\beta \mu)}{\beta,\beta {\bf h}} &
	-\thermodevin{n}{(\beta h_{\al})}{\beta,\beta \mu, \beta h_{\gamma}}&
	-\thermodevin{n}{\beta}{\beta \mu,\beta {\bf h}} \\
	\thermodevin{u_{\gamma}}{(\beta \mu)}{\beta,\beta {\bf h}} &
	-\thermodevin{u_{\gamma}}{(\beta h_{\al})}{\beta,\beta \mu, \beta h_{\gamma}}&
	-\thermodevin{u_{\gamma}}{\beta}{\beta \mu,\beta {\bf h}} \\
	\thermodevin{e}{(\beta \mu)}{\beta,\beta {\bf h}} &
	-\thermodevin{e}{(h_{\al})}{\beta,\beta \mu,\beta h_{\gamma}} &
	-\thermodevin{e}{\beta}{\beta \mu,\beta {\bf h}}
	\end{array}
	\right)\label{app_chi_macro_2a}\\
{\bm\chi_{nue}}(\bq)	&= \beta V\left(
	\begin{array}{ccc}
	\scal{\delta n^{*}(\bq) \delta n(\bq)}& \scal{\delta n^{*}(\bq) \delta u_{\gamma}(\bq)} & \scal{\delta n^{*}(\bq) \de(\bq)}  \\
	\scal{\delta u_{\al}^{*}(\bq) \delta n(\bq)}& \scal{\delta u_{\al}^{*}(\bq) \delta u_{\gamma}(\bq)} & \scal{\delta u_{\al}^{*}(\bq) \de(\bq)}\\
	\scal{\de^{*}(\bq) \delta n(\bq)} &\scal{\de^{*}(\bq) \delta u_{\gamma}(\bq)} & \scal{\de^{*}(\bq) \de(\bq)} 
	\end{array}
	\right) \label{app_chi_macro_2b}
\end{align}
\end{subequations}
\end{widetext}
Finally, in Eq.~\eqref{app_chi_macro_2b}, we have recovered the expression of the static susceptibility matrix $\bm{\chi_{nue}}$ we presented in Eq.~\eqref{app_chi_macro_1} from our consideration of the hydrodynamic variables in the Mori-Zwanzig formulation. In deriving Eq.~\ref{app_chi_macro_2}, from a purely thermodynamic starting point (Eq.~\eqref{app_Z}), we establish the relations (recall $h^{0}_{\albe}=0)$)
\begin{subequations}\label{app_conj_fs}
	\begin{align}
	&\da =\beta^{-1}\delta \left( \beta \mu \right) = \dmu - \frac{\mu^{0}}{T} \dT \\
	&\delta y_{\albe}=\beta^{-1}\delta (\beta h_{\albe})=\delta h_{\albe}, \\%\ (\because h^{0}_{\albe}=0)\\
	&\db = -\beta^{-1}\delta \beta = \frac{1}{T} \dT.
	\end{align}
\end{subequations}
Comparing Eq.~\eqref{app_mat_nue_conj_1} to Eq.~\eqref{app_conj_chi_nue} derived through the steps presented in Eq.~\eqref{app_tot_der_nue_1}, Eq.~\eqref{app_tot_der_nue_2} and Eq.~\eqref{app_chi_macro_2} allows us to identify the microscopically derived thermodynamic conjugate fields (see section~\ref{coarse_grain_react}) $\delta a, \delta y_{\albe}$ and $\delta b$ in terms of the thermodynamic intensive fields $\delta (\beta \mu),\delta (\beta h_{\albe})$ and $\delta \beta$ respectively. They are recalled in Eq.~\eqref{conj_fs} in the main text.

It is important to note here that the Eq.~\eqref{app_conj_fs} identifying $\delta a,\delta y_{\al}$ and $\delta b$ as intensive thermodynamic fields, can be derived %following a different albeit mathematically rigorous route 
also from considering the inverse route (see reference~\cite{Flo_thesis}). First reconsider the relation between $\delta a$, $\delta y_{\al}$, $\delta b$ and their respective conjugates $\delta n$, $\delta u_{\gamma}$, $\delta e$ through ${\bm\chi_{nue}^{-1}}$
in Eq.~\eqref{app_conj_chi_nue}.
%
% \begin{align}\label{conj_pairs_thermod}
%     &\begin{bmatrix}
%     \delta a  \\
%     \delta \bm{y}\\
%       \delta b
% \end{bmatrix}
%     =V\bm{\chi_{nue}}^{-1}
% \begin{bmatrix}
%   \delta n\\
%   \delta \bm{u}\\
%       \delta e
% \end{bmatrix}.
% \end{align}
%
After inverting the matrix ${\bm\chi_{nue}}$, using relations between partial derivatives of thermodynamic variables, Eq.~\eqref{app_chi_macro_2a} leads to
\begin{widetext}
\begin{subequations}\label{app_chi_inv_macro}
\begin{align}
	\lim_{\bq \rightarrow 0} {\bm\chi_{nue}}^{-1}(\bq)
	&	= (\beta V)^{-1} \left(
	\begin{array}{ccc}
	\thermodevin{(\beta \mu)}{n}{e,{\bf u}} &
	\thermodevin{(\beta \mu)}{u_{\gamma}}{n,u_{\al},e}&
	\thermodevin{(\beta \mu)}{e}{n,{\bf u}} \\
	-\thermodevin{(\beta h_{\al})}{n}{e,{\bf u}} &
	-\thermodevin{(\beta h_{\al})}{u_{\gamma}}{n,u_{\al},e}&
	-\thermodevin{(\beta h_{\al})}{e}{n,{\bf u}} \\
	-\thermodevin{\beta}{n}{e,{\bf u}} &
	-\thermodevin{\beta}{u_{\gamma}}{n,u_{\al},e}&
	-\thermodevin{\beta}{e}{n,{\bf u}}
	\end{array}
	\right),
	\end{align}
	\end{subequations}
\end{widetext}	
which can be plugged in Eq.~\eqref{rel_conjugates_cg} in order to bring out the expressions for the variables $\delta a$, $\delta b$ and $\delta y_{\albe}$ given in Eq.~\eqref{app_conj_fs}.

This representation of the ${\bm\chi_{nue}}^{-1}$ makes certain thermodynamic relations, derived in Eq.~\eqref{maxw_rel}, explicit. The expression for $C^{n}_{\albe\gamma\delta}$ (Eq.~\eqref{maxw_rel}c) is specially cited here because it helps us understand how a Voigt symmetric stiffness tensor $C^{n}_{\albe\gamma\delta}$ of dimensions $6\times 6$
\begin{align}\label{app_Cn_abcd}
    \beta C^{n}_{\albe\gamma\delta}q_{\be}q_{\delta}=\dfrac{\partial (\beta h_{\albe})}{\partial u_{\gamma\delta}}\Big|_{n,u_{\albe},\beta} q_{\be}q_{\delta}=-\dfrac{\partial (\beta h_{\al})}{\partial u_{\gamma}}\Big|_{n,u_{\al},\beta}
\end{align}
contributes to the $3\times 3$ (in a three dimensional system) block of correlations between the displacement fields in the ${\bm\chi_{nue}}^{-1}$ matrix. The results of this appendix discussing the relations between the coarse-grained fields derived from the Mori-Zwanzig projection operations and the corresponding thermodynamic fields, is used in section~\ref{therm_conj_macro}.

\section{Equation of motion : micro to macro}\label{eom_cg_app}
The equations of motion (Eq.~\eqref{EOM_micro_f}) for the microscopic relevant variables transform to the equations Eq.~\eqref{EOM_cg} when the fluctuations in the Bragg peak amplitudes $\delta n_{\bg}(\bq,t)$ are substituted with the two coarse-grained fields $\delta u_{\al}(\bq,t)$ and $\delta n(\bq,t)$ using the ansatz in Eq.~\eqref{ansatz1}. Here we present the steps involved in deriving Eq.~\eqref{EOM_cg} from Eq.~\eqref{EOM_micro_f}.
We reiterate the microscopic equations of motion (Eq.~\eqref{EOM_micro_f}) in the first lines of Eq.~\eqref{EOM_micro_f_app} %\com{now we need to mention that the $\Gamma^{j..}$ drop out in leading q-order. }
\begin{widetext}
\begin{subequations}\label{EOM_micro_f_app}
	\begin{align}
	\partial_{t}\delta n_{\bg}(\bq,t) &= \dfrac{\omega^{\rho j}_{\bg\al}(\bq)}{V} \delta \va(\bq,t) 
	- \sum_\bgp \frac{ \Gamma^{\rho\rho}_{\bg \bgp}(\bq)}{V}\dagp(\bq,t) -  \frac{\Gamma^{\rho e}_{\bg}(\bq)}{V} \db(\bq,t) \nonumber \\
	&=-i(g+q)_{\al}n_{\bg}\delta v_{\al}(\bq,t)-\sum_{\bg'}n_{\bg}n^{*}_{\bg'}g_{\al}g'_{\be}\zeta_{\albe}\delta a_{\bg'}(\bq,t)-q_{\be}\xi^\top_\albe n_{\bg} g_{\al}\delta b(\bq,t)\\
	\partial_{t}\delta e(\bq,t) &= \dfrac{\omega^{ej}_{\al}}{V} \delta \va(\bq,t) 
	- \sum_\bg \frac{\Gamma^{e\rho}_{\bg}(\bq)}{V}\daag(\bq,t) -\dfrac{\Gamma^{ee}}{V}\db(\bq,t)\nonumber\\
	&=-i(e_{0}+p_{0})q_{\al}\delta v_{\al}(\bq,t)-\sum_{\bg}n^{*}_{\bg}g_{\be}q_{\al}\xi_{\albe}\delta a_{\bg}(\bq,t)-q_{\al}q_{be}\alpha_{\albe}T\delta b(\bq,t)\\
	\partial_{t}\delta \ja(\bq,t) &= \sum_{\bg}\dfrac{\omega^{j\rho}_{\al\bg}}{V}\daag(\bq,t)
+\dfrac{\omega^{je}_{\al}}{V} \db(\bq,t) - \dfrac{\Gamma^{jj}_{\al\be}}{V} \delta v_{\be}(\bq,t)\nonumber\\
&=i\sum_{\bg}n^{*}_{\bg}(g+q)_{\al}\delta a_{\bg}(\bq,t)-i(e_{0}+p_{0})q_{\al}\delta b(\bq,t)-q_{\be}q_{\gamma}\eta_{\albe\gamma\delta}v_{\delta}(\bq,t)
	\end{align}
\end{subequations}
\end{widetext}
before plugging in expressions for $\omega^{\rho j}_{\bg\al}, \omega^{e j}_{\al}$ from Eq.~\eqref{freq_comp_micro} and $\Gamma^{\rho\rho}_{\bg\bg'}, \Gamma^{\rho e}_{\bg}, \Gamma^{jj}_{\albe}$ from Eq.~\eqref{ons_coef_set1}, Eq.~\eqref{ons_coef_set2} in the second lines of Eq.~\eqref{EOM_micro_f_app}. The components of $\bm{\Gamma}^{\rho j}_{(N\times 3)}$, $\bm{\Gamma}^{ e j}_{(1\times 3)}$ and their conjugate transposes in the dynamical equations  Eq.~\eqref{EOM_micro_f} have been neglected in our calculations. It can be shown~\cite{Flo_thesis} that the leading $\bq$ dependent term for these components arises from $\mathcal{O}\bq(\bg+\bq)$ while all the other components of $\bm{\Gamma}$ has $\mathcal{O}(\bg+\bq)$ (see Eq.~\eqref{ons_coef_set1}, Eq.~\eqref{ons_coef_set2} and Eq.~\eqref{ons_coef_set3}) leading order terms.

Next we differentiate Eq.~\eqref{ansatz3_sum}a with respect to time and substitute $\partial_{t}\delta n_{\bg}$ using Eq.~\eqref{EOM_micro_f_app}a. Then, due to the symmetry argument $\sum_g \lvert \ng \rvert^2 \gb=0$, several terms in Eq.~\eqref{EOM_micro_f_app}a vanishes and the time evolution of the density field is obtained

\begin{align}
    \partial_{t}\delta n(\bq,t)=-in_{0}q_{\al}\delta v_{\al}(\bq,t).
\end{align}
Similar steps are repeated, to get the time evolution of the displacement fields $\delta u_{\al}(\bq,t)$, starting with Eq.~\eqref{ansatz3_sum}b. It is differentiated with respect to time and then using Eq.~\eqref{EOM_micro_fa} and $\mathcal N_\albe = \sum_\bg \lvert \ng \rvert^2 \ga \gb$, one arrives at the following equation of motion for the displacement fields 
\begin{widetext}
\begin{subequations}\label{EOM_micro_macro_u_app}
\begin{align}
    \partial_{t}\delta u_{\al}(\bq,t)&=\delta v_{\al}(\bq,t)-\zeta_\albe\left[-\mub^*(\bq) \frac{\dn(\bq,t)}{n_0}+ \lambda_\bega(\bq) \duc(\bq,t) + \tau^*_\beta(\bq) \db(\bq,t)\right]- \ii \qb \xi_\albe^\top \db(\bq,t)\\
    &=\delta v_{\al}(\bq,t)-\zeta_{\albe}\delta y_{\be}- \ii \qb \xi_\albe^\top \db(\bq,t)
\end{align}
\end{subequations}
\end{widetext}
Here we have made use of the fact that $\delta a_{\bg}$ can be written in terms of the inverse density correlation matrix $J_{\bg\bg'}$ (see Eq.~\eqref{rel_conjugates3}a) which ends up giving the expressions for the generalised material coefficients $\lambda_{\albe},\mu_{\al},\nu,\tau_{\al},\theta$ (see Appendix~\ref{mat_coef_app}). For a more elegant representation of Eq.~\eqref{EOM_micro_macro_u_app}a, we substitute the second term with $\delta y_{\be}$ from Eq.~\eqref{conj_coarse-grained}c.

The equation for the energy density (Eq.~\eqref{EOM_micro_f_app}b) is taken up next. Once again using the definitions of the generalised material coefficients (see Appendix~\ref{mat_coef_app}) and the expression for $\delta y_{\be}$ in Eq.~\eqref{conj_coarse-grained}c leads to the following equations
\begin{widetext}
\begin{subequations}\label{EOM_micro_macro_e_app}
	\begin{align}
	\partial_{t}\delta e(\bq,t) &=
	 -\ii \left(e_0 +p_0 \right) \qa \delta v_{\al}(\bq,t) 
	 +\ii \qa \xi_\albe \left[-\mub^*(\bq) \frac{\dn(\bq,t)}{n_0}  +  \lambda_\bega(\bq) \uc(\bq,t)+  \tau_\beta^*(\bq) \db(\bq,t)\right]
	 - \qa \qb \alpha_\albe T \db(\bq,t)\\
	 &=
	 -\ii \left(e_0 +p_0 \right) \qa \delta v_{\al}(\bq,t) 
	 +\ii \qa \xi_\albe \delta y_{\be}
	 - \qa \qb \alpha_\albe T \db(\bq,t).
	\end{align}
\end{subequations}
\end{widetext}
Finally, the equation for the density of the linear momentum in Eq.~\eqref{EOM_micro_f_app}c can be written in terms of the conjugate fields $\delta a,\delta b,\delta v_{\al}, \delta y_{\al}$ defined in Eq.~\eqref{conj_coarse-grained}. Here we have employed the same mathematical manipulations, as in case of the equations for $\delta n, \delta u_{\al}, \delta e$, to identify the generalised material constants and their relations to the conjugate fields.
\begin{widetext}
\begin{subequations}\label{EOM_micro_macro_j_app}
\begin{align}
\partial_{t}\delta \ja(\bq,t) &= -\ii n_{0}\qa\left[ \dfrac{\nu(\bq)}{n^{2}_{0}} \dn(\bq,t) - \dfrac{\mu_\beta(\bq) }{n_{0}}\dub(\bq,t)  - \dfrac{\theta^*(\bq)}{n_{0}}\db(\bq,t)\right]\nonumber\\
&\ghostequal -\left[- \mu^*_\alpha(\bq) \frac{\dn(\bq,t)}{n_0}+ \lambda_\albe(\bq) \dub(\bq,t) +\tau_\alpha^*(\bq) \db(\bq,t) \right]\nonumber \\
&\ghostequal- \ii \left(e_0 +p_0 \right) \qa \db(\bq,t) - \qb \qc \eta_{\albe \gade} \vd(\bq,t)\\
&=-in_{0}q_{\al}\delta a(\bq,t)-\delta y_{\al}(\bq,t)
- \ii \left(e_0 +p_0 \right) \qa \db(\bq,t) - \qb \qc \eta_{\albe \gade} \vd(\bq,t).
\end{align}
\end{subequations}
\end{widetext}

This appendix shows how, the eight hydrodynamic equations presented in Eq.~\eqref{EOM_cg} can be derived from the $(N+4)$ Mori-Zwanzig equations (Eq.~\eqref{EOM_micro_f}) for the microscopic fields of a local-defect rich three dimensional crystal, through the coarse-graining ansatz in Eq.~\eqref{ansatz1}.

\bibliography{sg_bib}% Produces the bibliography via BibTeX.

\end{document}